\documentclass[twocolumn,fleqn,usenatbib,useAMS]{aastex63}
\usepackage{newtxtext,newtxmath}
\usepackage[T1]{fontenc}
\usepackage{ae,aecompl}
\usepackage{graphicx}
\usepackage{natbib}
\usepackage{amssymb}
\usepackage{amsmath}
\usepackage{xcolor}

\def\lesssim{\lower.5ex\hbox{$\; \buildrel < \over \sim \;$}}
\def\gtrsim{\lower.5ex\hbox{$\; \buildrel > \over \sim \;$}}

\received{}
\revised{}
\accepted{}

\submitjournal{ApJ}

\shorttitle{SAMs on FIRE}
\shortauthors{Pandya and the SMAUG Collaboration}

\begin{document}

\title{First results from SMAUG: The need for preventative stellar feedback and improved baryon cycling in semi-analytic models of galaxy formation}

\correspondingauthor{Viraj Pandya}
\email{viraj.pandya@ucsc.edu}

\author[0000-0002-2499-9205]{Viraj Pandya}
\affiliation{UCO/Lick Observatory, Department of Astronomy and Astrophysics, University of California, Santa Cruz, CA 95064, USA}
\affiliation{Center for Computational Astrophysics, Flatiron Institute, New York, NY 10010, USA}

\author{Rachel S. Somerville}
\affiliation{Center for Computational Astrophysics, Flatiron Institute, New York, NY 10010, USA}
\affiliation{Department of Physics and Astronomy, Rutgers University, 136 Frelinghuysen Rd, Piscataway, NJ 08854, USA}

\author{Daniel Angl\'es-Alc\'azar}
\affiliation{Center for Computational Astrophysics, Flatiron Institute, New York, NY 10010, USA}
\affiliation{Department of Physics, University of Connecticut, 196 Auditorium Road, U-3046, Storrs, CT 06269-3046, USA}

\author{Christopher C. Hayward}
\affiliation{Center for Computational Astrophysics, Flatiron Institute, New York, NY 10010, USA}

\author{Greg L. Bryan}
\affiliation{Center for Computational Astrophysics, Flatiron Institute, New York, NY 10010, USA}
\affiliation{Department of Astronomy, Columbia University, 550 West 120th Street, New York, NY 10027, USA}

\author{Drummond B. Fielding}
\affiliation{Center for Computational Astrophysics, Flatiron Institute, New York, NY 10010, USA}

\author{John C. Forbes}
\affiliation{Center for Computational Astrophysics, Flatiron Institute, New York, NY 10010, USA}

\author{Blakesley Burkhart}
\affiliation{Center for Computational Astrophysics, Flatiron Institute, New York, NY 10010, USA}
\affiliation{Department of Physics and Astronomy, Rutgers University, 136 Frelinghuysen Rd, Piscataway, NJ 08854, USA}

\author{Shy Genel}
\affiliation{Center for Computational Astrophysics, Flatiron Institute, New York, NY 10010, USA}
\affiliation{Columbia Astrophysics Laboratory, Columbia University, 550 West 120th Street, New York, NY 10027, USA}

\author{Lars Hernquist}
\affiliation{Harvard-Smithsonian Center for Astrophysics, 60 Garden Street, Cambridge, MA 02138, USA}

\author[0000-0003-2896-3725]{Chang-Goo Kim}
\affiliation{Center for Computational Astrophysics, Flatiron Institute, New York, NY 10010, USA}
\affiliation{Department of Astrophysical Sciences, Princeton University, Princeton, NJ 08544, USA}

\author{Stephanie Tonnesen}
\affiliation{Center for Computational Astrophysics, Flatiron Institute, New York, NY 10010, USA}

\author{Tjitske Starkenburg}
\affiliation{Center for Computational Astrophysics, Flatiron Institute, New York, NY 10010, USA}
\affiliation{CIERA and Department of Physics and Astronomy, Northwestern University, 2145 Sheridan Road, Evanston, IL 60208, USA}

\begin{abstract}
Semi-analytic models (SAMs) are a promising means of tracking the physical processes associated with galaxy formation, but many of their approximations have not been rigorously tested. As part of the SMAUG (Simulating Multiscale Astrophysics to Understand Galaxies) project, we compare predictions from the FIRE-2 hydrodynamical ``zoom-in" simulations to those from the Santa Cruz SAM run on the same halo merger trees, with an emphasis on the global mass flow cycle. Our study includes 13 halos spanning low-mass dwarfs ($M_{\rm vir}\sim10^{10}M_{\odot}$ at $z=0$), intermediate-mass dwarfs ($M_{\rm vir}\sim10^{11}M_{\odot}$) and Milky Way-mass galaxies ($M_{\rm vir}\sim10^{12}M_{\odot}$). The SAM and FIRE-2 predictions agree relatively well with each other in terms of stellar and ISM mass, but differ dramatically on CGM mass (the SAM is lower than FIRE-2 by $\sim3$ orders of magnitude for dwarfs). Strikingly, the SAM predicts higher gas accretion rates for dwarfs compared to FIRE-2 by factors of $\sim10-100$, and this is compensated for with higher mass outflow rates in the SAM. We argue that the most severe model discrepancies are caused by the lack of preventative stellar feedback and the assumptions for halo gas cooling and recycling in the SAM. As a first step towards resolving these model tensions, we present a simple yet promising new preventative stellar feedback model in which the energy carried by supernova-driven winds is allowed to heat some fraction of gas outside of halos to at least the virial temperature such that accretion is suppressed.
\end{abstract}


\section{Introduction}
In the $\Lambda$CDM paradigm of galaxy formation, the growth of dark matter halos is paralleled by the accretion of gas from the intergalactic medium \citep[IGM; e.g.,][]{white78,blumenthal84}. The accreted gas is thought to reside within the circumgalactic medium (CGM), which acts as a buffer between the interstellar medium (ISM) and the IGM. Radiative cooling of this CGM gas leads to the build-up of the ISM and eventually star formation. The resulting feedback from stars and supernovae is capable of heating and ejecting gas from the ISM back into the CGM or IGM, and the energy and momentum carried by these stellar-driven winds can also suppress future gas cooling and accretion (and hence star formation). Gas that has been previously ejected from the ISM can re-accrete, which together with the other gas flow processes gives rise to the ``baryon cycle" of galaxies \citep[e.g.,][]{oppenheimer10,christensen16,anglesalcazar17}. These and other physical processes ultimately shape the evolutionary histories of individual galaxies, with the statistical properties of galaxy populations (e.g., the stellar mass function and galaxy scaling relations) emerging as a result. This is the modern high-level picture of galaxy formation gleaned from both observations and interpretive models, but many uncertainties remain in our detailed understanding of the relevant physics \citep[see the recent reviews by][]{somervilledave15,naab17}.

Models of galaxy formation span a continuum in terms of volume and resolution. To thoroughly understand galaxy formation in a cosmological context requires modeling populations of galaxies, which in turn requires modeling large volumes (several 100$^3$ Mpc$^3$). Such large-volume population studies are important to: (1) ensure a robust sampling of the scatter in halo growth histories at a fixed mass, (2) explore the range of physical processes at play across different large-scale environments, (3) enable comparisons to observations from large-volume surveys, and (4) allow galaxies to ultimately be used as reliable cosmological probes. However, owing to resolution limits, all currently existing large-volume models contain a ``phenomenological" component, which is to say that: (1) physics occurring below the resolution limit is parameterized, often in an ad hoc way, and (2) the free parameters of the model are adjusted to match a limited set of observations. This is generally true for modern large-volume hydrodynamical simulations, which solve the equations of gravity and fluid dynamics along with ``subgrid recipes" \citep[e.g.,][and references therein]{vogelsberger14,genel14,schaye15,pillepich18,dave19}. It is also true for semi-analytic models (SAMs), which attempt to distill the key insights from more sophisticated simulations using a set of coupled ordinary differential equations that track the flow of mass between different galactic components \citep[e.g.,][]{whitefrenk91,kauffmann93,cole94,somerville99}. Both phenomenological approaches have their advantages, disadvantages and simplifying assumptions, but ultimately they are complementary and inform each other.

There is a long history of comparing the predictions of SAMs to hydrodynamical simulations. \citet{benson01} first demonstrated how the parameters controlling halo gas cooling and merger rates in a simplified SAM could be adjusted to better match predictions from a cosmological hydrodynamical simulation. They focused on the cosmic number and mass densities of hot halo gas and dense ISM gas. Overall, their study showed remarkable consistency between the two very different approaches for modeling halo gas cooling in cosmological volumes (in an average statistical sense). Subsequently, \citet{yoshida02} and \citet{helly03} each ran their own simplified SAM on halo merger trees extracted directly from hydrodynamical simulations and compared predictions for gas cooling and accretion on an individual halo-by-halo basis. Both of these studies demonstrated the striking correspondence, with minimal systematic offsets, between their SAM and hydrodynamical predictions. In the years since, there have been a number of studies that compared the predictions of SAMs and hydrodynamical simulations (using both statistical and individual halo-by-halo approaches). Owing to the ever-increasing sophistication of the simulations, the comparisons have expanded to include a wider range of physical processes beyond just halo gas cooling: evolution of dark matter subhalos \citep{jiang16}, UV background heating due to spatiotemporally inhomogeneous reionization \citep{mutch16}, cold/rapid versus hot/slow mode accretion \citep{cattaneo07,lu11,hirschmann12}, relating halo and galaxy angular momentum \citep{guo16,stevens17,mitchell18}, multi-phase ISM and dust modeling \citep[e.g.,][]{popping19}, feedback processes \citep{weinmann12,qin18,ayromlou20} and baryonic effects on dark matter halo concentrations \citep[e.g.,][]{dutton16}. 

Among the many SAM versus hydrodynamical simulation comparisons, the studies by \citet{stringer10} and \citet{neistein12} are particularly informative. \citet{stringer10} modified several aspects of an existing SAM to ask how closely it could reproduce the evolution of a single Milky Way (MW)-mass halo simulated at high resolution. They found remarkable potential in the ability of their SAM to match the predictions of the more sophisticated simulation as a function of time, including the evolution of shocked versus unshocked halo gas accretion, halo gas scale length, disk gas scale length, disk circular velocity, stellar mass, cold gas mass, hot gas mass, hot disk gas mass and outflow gas mass. They further showed that their fiducial, previously published SAM (used for observational comparisons) predicted a very different evolution for the same simulated MW-mass halo, primarily due to its assumptions of much lower star formation efficiency and much stronger supernova feedback. \citet{neistein12} went a step further and characterized the efficiencies of various processes in a large-volume hydrodynamical simulation using a novel particle phase tracking approach. They derived mass- and redshift-dependent functions that summarized accretion, cooling, star formation and feedback in the simulation. They emphasized that these functions were significantly different than the assumptions built into traditional SAMs, but that the functions represented a common language for connecting SAMs and simulations.

It is clear from the many previous studies that SAMs show the potential to transparently summarize the complicated physics of and emergent predictions from more sophisticated cosmological hydrodynamical simulations. However, an outstanding question that still remains today is whether modifications made to SAMs to bring them into better agreement with simulations must also necessarily come at the expense of no longer matching observations \citep{cattaneo07,stringer10}. Primarily, this puzzle must be driven by the fact that SAMs include only a limited description of the full range of phenomena found in simulations. However, a secondary cause is the somewhat circular logic of comparing to reference simulations that are themselves phenomenologically calibrated and hence effectively SAM-like in nature \citep[e.g.,][]{crain15,pillepich18}. Even if such simulations agree with a plethora of observations, the choice of subgrid model and associated free parameters carry degeneracies that propagate as largely unknown systematic uncertainties on predictions for galaxy populations. These uncertainties make it difficult to firmly interpret observations, but this has motivated important recent work on improving the flexibility and computational efficiency of SAMs \citep[e.g.,][]{henriques09,lu11b,henriques13,lagos18,forbes19}. 

Given the ambiguities associated with comparing phenomenological models, it is also interesting to compare the predictions of SAMs to higher resolution cosmological ``zoom-in" simulations where small-scale physical processes (namely, stellar feedback) are implemented locally and more self-consistently. Here we focus on the Feedback In Realistic Environments (FIRE) Project\footnote{http://fire.northwestern.edu} \citep[the second generation FIRE-2 suite;][]{hopkins14,hopkins18} and the Santa Cruz SAM \citep[the most recent version:][]{somerville15}, both of which have been shown to reproduce a large range of observations. The FIRE-2 simulations represent a good comparison suite because their stellar feedback model deposits mass, energy, momentum and metals locally without any explicit ``tuning" to match observations; the resulting large-scale effects are hence emergent phenomena \citep[e.g., outflows, CGM heating and recycling;][]{muratov15,anglesalcazar17}. As with any simulation, caution is warranted regarding the absolute correctness and completeness of the FIRE-2 simulations (improvements can always be made to the numerical algorithms, the range of physical processes implemented, and the diversity of halo mass accretion histories and large-scale environments probed). However, for the purposes of improving physical prescriptions for SAMs, we can confidently use FIRE-2 as a baseline for comparison, identify systematic discrepancies, and develop plausible solutions to guide future work. With the FIRE-2 suite, we will study the time evolution of 13 individual halos across a broad range in mass: low-mass dwarfs ($M_{\rm vir}\sim10^{10}M_{\odot}$ at $z=0$), intermediate-mass dwarfs ($\sim10^{11}M_{\odot}$), and MW-mass galaxies ($\sim10^{12}M_{\odot}$). We will also restrict the scope of our comparison to a few bulk quantities that characterize the overall baryon cycle of galaxies (the foundation of any SAM): stellar, ISM and CGM masses, and the corresponding mass inflow and outflow rates for the ISM and CGM. The inclusion of flow \emph{rates} in addition to global bulk quantities is, to our knowledge, a novel feature of this work which has not been widely studied in the past \citep[but see][]{hirschmann12}.

This paper advances one of the key goals of the SMAUG Collaboration\footnote{Simulating Multiscale Astrophysics to Understand Galaxies} which is to ask: is it possible to develop a model that faithfully captures the essential physics of galaxy formation in a more computationally efficient way than fully numerical large-volume simulations? Given that the physical processes involved in galaxy formation are not fully understood and also span a vast range in scale, it is not feasible to develop a single ``ab initio" simulation that is capable of making credible predictions on the scale of galaxy populations. Instead, SMAUG aims to carefully design a suite of high-resolution numerical experiments whose results can be coarse grained to develop realistic subgrid prescriptions for cosmological simulations. As part of the first results from SMAUG\footnote{https://www.simonsfoundation.org/flatiron/center-for-computational-astrophysics/galaxy-formation/smaug/papersplash1}, the resolved ISM simulations by \citet{kim20} and resolved black hole accretion simulations by \citet{anglesalcazar20} take the first step towards this goal. The complementary parameter space study of simulated star-forming regions by \citet{motwani20} is designed to provide the initial conditions for a future suite of resolved ISM simulations building on \citet{kim20}. In the present work, we take the first step towards re-tooling and calibrating SAMs using physically self-consistent simulations instead of observations so that SAMs may become more predictive rather than descriptive in nature. Our emphasis on the need to improve phenomenological modeling of stellar feedback and gas flows in the CGM underscores the work of \citet{fielding20}, who find that the properties of the multi-phase CGM depend strongly on the nature of feedback, cosmological accretion and simulation methodology. 

This paper is organized as follows. We describe the FIRE-2 simulations and the Santa Cruz SAM in \autoref{sec:models}, and our analysis methods in \autoref{sec:analysis}. We present the results of our comparison in \autoref{sec:results}, while \autoref{sec:discussion} is devoted to interpreting the model discrepancies and presenting a preventative stellar feedback model for inclusion in future SAMs. A summary is provided in \autoref{sec:summary}. In Appendix \ref{sec:dmonly}, we compare predictions from the SAM run on merger trees extracted from the hydrodynamical simulations versus corresponding dark matter (DM) only simulations.  

\section{Model Descriptions}\label{sec:models}
Here we describe the FIRE-2 cosmological hydrodynamical ``zoom-in" simulations and the Santa Cruz SAM. Note that the FIRE-2 simulations assume $h=0.70$, $\Omega_{m,0}=0.27$, $\Omega_{\Lambda,0}=0.73$ and $\Omega_b=0.045$ \citep[see section 2.8 of][]{hopkins18}. The Santa Cruz SAM assumes the \citet{planck16} cosmology, with $h=0.678$, $\Omega_{m,0}=0.308$, $\Omega_{\Lambda,0}=0.692$ and $\Omega_b=0.0486$. The main differences between the SAM and FIRE-2 that we focus on in this paper are unlikely to be driven by the small differences in assumed cosmology.

\subsection{FIRE-2 Simulations}
We use the FIRE-2 suite of cosmological hydrodynamical ``zoom-in" simulations described in \citet{hopkins18}. The simulations were run with the Gizmo\footnote{http://www.tapir.caltech.edu/~phopkins/Site/GIZMO.html} code \citep{hopkins15} using the Lagrangian meshless finite-mass method and fully-adaptive gravitational force softening lengths for gas. Briefly, a large DM-only box was evolved to $z=0$, and relatively isolated halos were chosen to be re-simulated at much higher resolution with baryons included. The initial zoom region is defined to be $\sim5R_{\rm vir}$ around the halo at $z=0$, but in practice only the zoom region within $\sim2R_{\rm vir}$ is guaranteed to avoid contamination from low resolution DM particles.

The FIRE-2 simulations account for gas heating and cooling between temperatures of $10$K and $10^{10}$K, including free-free, photoionization/recombination, Compton, photoelectric, metal-line, molecular, fine-structure, dust collisional and cosmic ray processes \citep[the corresponding cooling tables are given in Appendix B of][]{hopkins18}. A spatially uniform but redshift dependent UV background is imposed based on \citet{fauchergiguere09}. Star formation occurs stochastically in self-gravitating, molecular, self-shielding gas that has hydrogen number density $n_H\geq1000$ cm$^{-3}$. Owing to the high spatial and mass resolution (see below), stellar feedback is modeled via local deposition of mass, momentum, energy and metal mass from star particles to neighboring gas particles. The feedback accounts for both Type Ia and Type II supernovae, stellar winds, momentum from radiation pressure, photo-ionization and photo-electric heating. In this way, the generation, propagation and recycling of large-scale galactic winds are emergent phenomena rather than being put in ``by hand" via delayed cooling, thermal bombs or decoupled winds \citep[e.g.,][]{muratov15,anglesalcazar17}. 

Of the 27 high resolution FIRE-2 halos listed in Table 1 of \citet{hopkins18}, we use the 13 halos for which particle data were output for the full set of 600 snapshots: m10q, m10v, m10y, m10z, m11a, m11b, m11q, m11c, m11v, m11f, m12i, m12f and m12m \citep[these specific halos were originally presented in][]{wetzel16,garrisonkimmel17,chan18,hopkins18}. Our sample of FIRE-2 halos is identical to those of \citet[][see their Table 1]{hafen19}, with the addition of m11f but excluding their metal diffusion runs. The halos are grouped into three mass bins based on their $z=0$ virial mass: the four m10 halos have $M_{\rm vir}\sim10^{10}M_{\odot}$ (low-mass dwarf bin), the six m11 halos have $M_{\rm vir}\sim10^{11}M_{\odot}$ (intermediate-mass dwarf bin), and the three m12 halos have $M_{\rm vir}\sim10^{12}M_{\odot}$ (MW-mass halo bin). With this sample, we will be able to study systematic trends with halo mass for discrepancies between the SAM and FIRE-2. The mass and spatial (gravitational force softening) resolution vary with halo mass, and are systematically higher for the dwarfs. The star/gas particle masses are $250M_{\odot}$ for the m10 halos, $880M_{\odot}$ for m11q, $2100M_{\odot}$ for m11a, m11b and m11c, $7100M_{\odot}$ for m11v, m12i, m12f and m12m, and $12000M_{\odot}$ for m11f. The DM particle masses are roughly $\sim5\times$ higher. The minimum adaptive gravitational force softening lengths for the gas are on the order of $\sim1$ pc \citep[see][for more details]{hopkins18}. In addition, the typical snapshot spacing is $\sim20$ Myr, which allows us to accurately track variability in halo mass accretion and star formation for comparison to the SAM.

\subsection{Santa Cruz Semi-Analytic Model}\label{sec:scsam}
The Santa Cruz SAM was first presented in \citet{somerville99}, with significant updates described in \citet{somerville08}, \citet{somerville12}, \citet{porter14}, \citet{popping14} and \citet{somerville15}. Here, we use the latest \citet{somerville15} version, which includes recipes for multi-phase partitioning of ISM gas. We adopt the same calibration of free parameters for this version as used in \citet{popping19}; we will report the adopted parameter values for each of the relevant equations that we review below. We will not review the details of satellite-specific processes since our comparison to FIRE-2 only involves central halos \citep[section 2.8 and section 2.1, respectively, of][describe how subhalos are modeled in the SAM]{somerville99,somerville08}. In addition, since active galactic nucleus (AGN) feedback is not implemented in the FIRE-2 simulations employed here, we have disabled it in the SAM for a more consistent comparison and will not review the corresponding equations here.\footnote{Note that AGN feedback can have appreciable effects for MW halos but not dwarfs in the SAM \citep{somerville08}. Nevertheless, for a consistent comparison with the FIRE-2 simulations we must disable it in the SAM.} 

We emphasize that we have not made any other changes to the Santa Cruz SAM used in previously published works. The initial mass function assumed in the SAM \citep{chabrier03} and FIRE-2 \citep{kroupa01} are similar enough that they are unlikely to drive any significant model differences. For the cooling function, the SAM assumes \citet{sutherland93} whereas FIRE-2 has an implicit cooling algorithm based on more recent calculations for a wide range of physical processes. It is possible that if we implement the FIRE-2 cooling function approximations listed in Appendix B of \citet{hopkins18}, the SAM predictions for CGM cooling rates could change dramatically. However, we do not think the different cooling functions can explain the more fundamental qualitative differences between the two models (e.g., regarding halo gas accretion and recycling). On the other hand, the metallicity modeling and calibration are quite different between the SAM and FIRE-2, with the former using the instantaneous recycling approximation and only considering Type II supernovae with an assumed chemical yield $y=1.6$ \citep[section 2.8 of][]{somerville08}. In contrast, FIRE-2 self-consistently tracks chemical yields of different species during various stages of star particle evolution \citep[Appendix A of][]{hopkins18}. Although we do not focus on comparing metallicities between the two models in this paper, a more sophisticated treatment of metals in the SAM could affect cooling-related processes and have observable consequences (we defer an investigation of this to future work).

\subsubsection{Halo gas accretion}
For any given halo, the SAM begins by computing the DM accretion rate via finite differencing of the virial mass time series provided by the halo merger tree. Before the Universe is reionized (reionization is assumed to occur instantaneously at a specified redshift), it is assumed that gas accretion into the halo perfectly tracks DM accretion with the universal baryon fraction, i.e., $\dot{M}_{\rm gas}=f_{\rm b}\dot{M}_{\rm vir}$ where $f_{\rm b}=0.158$ according to \citet{planck16}. After reionization, the pristine gas accretion rate is suppressed due to photoheating from the UV background: 
\begin{equation}\label{eqn:pristine}
\dot{M}_{\rm CGM,in,pristine}=f_{\rm coll}f_{\rm b}\dot{M}_{\rm vir}.
\end{equation}
The factor $f_{\rm coll}$ gives the fraction of infalling baryonic mass that is able to collapse into the halo despite heating by the UV background. It depends on halo mass and redshift, and is taken from \citet{okamoto08} who characterized the suppressive effects of the \citet{haardt01} UV background in their hydrodynamical simulations. In practice, the formula for $f_{\rm coll}$ involves computing a ``characteristic filtering mass" at which the gas accretion rate drops to half of the universal $f_{\rm b}$; above this characteristic halo mass, the accretion rate approaches $f_{\rm b}$, and below it the accretion rate drops steeply such that UV background heating is more effective in lower mass halos. The filtering halo mass is computed according to Appendix B of \citet{kravtsov04}; it is $M_{\rm filt}\lesssim10^8M_{\odot}$ before reionization is complete, and rises to $M_{\rm filt}\approx10^{10}M_{\odot}$ by $z=0$. We assume the IGM is fully reionized by $z\sim8$, consistent with \citet{planck16}.  All of our FIRE-2 halos have virial masses above the characteristic filtering mass at all times $z\lesssim10$, except for m10v which becomes larger starting only at $z\sim2$ \citep[see also Figure 11 of][]{fitts17}. We have experimented with changing the filtering mass normalization to mimic using different UV background models, and find that our results are insensitive for reasonable changes.

On top of the pristine IGM gas accretion, the SAM adds the ``re-accretion'' of gas that was previously ejected from the halo due to stellar feedback: 
\begin{equation}
\dot{M}_{\rm CGM,in,recycled} = \chi_{\rm re-infall} \left( \frac{M_{\rm ejected}}{t_{\rm dyn}} \right).
\end{equation}
$M_{\rm ejected}$ is the total mass of the ejected gas reservoir (its growth rate is set by the product of \autoref{eqn:mdot_out_ism} and \autoref{eqn:feject} described below) and $\chi_{\rm re-infall}$ is a free parameter that sets what fraction of the ejected gas reservoir can cool back into the halo at each time step. We assume $\chi_{\rm re-infall}=0.1$ as in previous Santa Cruz SAM papers; this implies that the ejected gas will re-accrete back into the halo on ten dynamical times $t_{\rm dyn}\equiv\frac{R_{\rm vir}}{V_{\rm vir}}$, where $V_{\rm vir}=\sqrt{\frac{GM_{\rm vir}}{R_{\rm vir}}}$ is the circular velocity of the halo at the virial radius (note that $t_{\rm dyn}\approx0.1 t_{\rm Hubble}$, so the gas will effectively re-accrete over a Hubble time).

There are two additional sources of CGM gas from within the halo itself. The first is outflows from the ISM that get deposited into the CGM; we defer this to the discussion of the relevant stellar feedback equations below. The second source is transfer from subhalos: the SAM assumes that once a halo becomes a subhalo, the CGM of the subhalo is instantaneously transferred to the CGM of the host halo. Although physical processes associated with satellite galaxies (i.e., subhalos) can indirectly affect the evolution of the central galaxy, we do not expect these processes to be the dominant ones in the simulations we are considering.

\subsubsection{CGM gas cooling}\label{sec:coolingmodel}
Gas that has accreted into the halo as described above builds up the CGM mass. The cooling rate of this CGM gas into the ISM is computed according to \citet{whitefrenk91}, which is also the basis for most, if not all, other SAMs. First, the CGM is assumed to uniformly be at the virial temperature of the halo at each time step. Then, the ``radiative cooling time" is computed, which is the characteristic timescale for the gas to cool by radiating away its current thermal energy: 
\begin{equation}\label{eqn:tcool}
t_{\rm cool}(r) = \frac{(3/2)\mu m_p kT_{\rm vir}}{\rho_g(r) \Lambda(T_{\rm vir},Z_h)}.
\end{equation}
$\mu m_p$ is the mean molecular weight of the halo gas and $\Lambda(T_{\rm vir},Z_h)$ is the \citet{sutherland93} cooling function, which takes into account the metallicity of the halo gas $Z_h$. As is common practice, the gas mass density radial profile is assumed to be a singular isothermal sphere: $\rho_g(r) = \frac{M_{\rm CGM}}{4\pi R_{\rm vir} r^2}$. Plugging this into the equation for $t_{\rm cool}$, one can solve for $R_{\rm cool}$, the radius within which all of the gas can radiatively cool within $t_{\rm cool}$ (heating is neglected). Then, integrating to compute the total cooled mass within $R_{\rm cool}$ and differentiating with respect to time gives the ISM mass accretion rate: 
\begin{equation}
\dot{M}_{\rm ISM,in} = \frac{1}{2} M_{\rm CGM} \frac{R_{\rm cool}}{R_{\rm vir}} \frac{1}{t_{\rm cool}}.
\end{equation}
Note that although different choices have been adopted in the literature, it is common practice to assume that the cooling time is equal to the halo dynamical time at $R_{\rm vir}$, i.e., $t_{\rm cool}=t_{\rm dyn}$. It is possible to have $R_{\rm cool}>R_{\rm vir}$ (this generally occurs for low mass halos), and these instances are assumed to represent ``cold/fast/filamentary" mode accretion. Since $R_{\rm cool}>R_{\rm vir}$ implies that the cooling time is shorter than the dynamical time, the SAM ignores the radiative cooling prediction during these timesteps and instead sets the ISM accretion rate equal to the halo gas accretion rate \citep[see also, e.g.,][]{croton06}. Otherwise, the interpretation is that gas has been gravitationally shock-heated to the virial temperature upon first accreting into the halo, and is now radiatively cooling via the assumed ``hot/slow/spherical" mode. As mentioned in \citet{somerville08}, reasonable variations within the framework of this particular gas cooling model (e.g., changing the definition of $t_{\rm cool}$ or assuming a different form for $\rho_g(r)$) can lead to variations in the ISM accretion rate by a factor of at most $\sim2-3$. 

\subsubsection{Star formation and stellar feedback}\label{sec:sf}
Gas that has accreted into the ISM is partitioned into HI, H$_2$, HII and metals \citep[details given in][]{popping14,somerville15}. The SAM keeps track of the mass surface density for these different gas phases in radial disk annuli \citep[see][for details about the SAM disk model]{somerville08b}. The default recipe for predicting the star formation rate (SFR) surface density is based on the molecular hydrogen gas phase alone, accounting for a higher conversion efficiency above a critical H$_2$ surface density \citep{bigiel08,bigiel11,narayanan12}:
\begin{equation}
\Sigma_{\rm SFR} = A_{\rm SF} \left( \frac{\Sigma_{\rm H_2}}{10 M_{\odot} \rm pc^{-2}} \right) \left( 1 + \frac{\Sigma_{\rm H_2}}{\Sigma_{\rm H_2,crit}} \right)^{N_{\rm SF}} . 
\end{equation}
$A_{\rm SF}$, $N_{\rm SF}$ and $\Sigma_{\rm H_2,crit}$ are free parameters of this two-part scaling relation. We assume $A_{\rm SF}=5.98\times10^{-3}$ M$_{\odot}$ yr$^{-1}$ kpc$^{-2}$, $N_{\rm SF}=1.0$ and  $\Sigma_{\rm H_2,crit}=70$ M$_{\odot}$ pc$^{-2}$ \citep[following][]{popping14,popping19}. There are various ways to estimate the molecular hydrogen gas density $\Sigma_{\rm H_2}$. Here we use the metallicity-dependent partitioning approach of \citet{gnedin11} that is the default in the Santa Cruz SAM.

On top of the continuous ``disk mode" star formation, the SAM also superimposes ``starbursts" due to galaxy mergers. The SFR spikes are modeled using a Laplace distribution (i.e., double exponential distribution) whose two parameters, the total starburst mass $M_{\rm burst}$ and the associated timescale $\tau_{\rm burst}$, are a function of progenitor properties and calibrated to binary galaxy merger simulations \citep[][and references therein]{somerville08,porter14}. Note that while starbursts will contribute some variability to the overall star formation history (SFH), the disk star formation can exhibit its own stochasticity due to changes in the H$_2$ gas fraction (driven by changes in gas metallicity and galaxy size) and changes in the overall gas fraction (driven by stellar feedback and CGM gas cooling).

All stellar feedback in the SAM (aside from heating by the UV background) is ejective. At every timestep, the mass outflow rate from the ISM due to stellar feedback is computed as:
\begin{equation}\label{eqn:mdot_out_ism}
\dot{M}_{\rm ISM,out} = \epsilon_{\rm SN} \left(\frac{V_{\rm max}}{V_0}\right)^{\alpha_{\rm SN}} \dot{M}_{\rm SFR}.
\end{equation}
Here, $\epsilon_{\rm SN}$ and $\alpha_{\rm SN}$ are free parameters, $V_0$ is an arbitrary normalization constant and $V_{\rm max}$ is the maximum circular velocity of the halo taken from the merger tree. We assume $\epsilon_{\rm SN}=1.5$ and $\alpha_{\rm SN}=-2.6$ following \citet[][]{popping19}. The total mass blown out of the ISM is either transferred into the CGM or driven out of the halo completely (i.e., deposited into the ejected reservoir). The fraction of outflow mass that gets ejected from the halo is computed via 
\begin{equation}\label{eqn:feject}
f_{\rm eject} = [1.0 + (V_{\rm vir} / V_{\rm eject})^{\rm \alpha_{\rm eject}}]^{-1}\;,
\end{equation}
where $\alpha_{\rm eject}$ and $V_{\rm eject}$ are free parameters, with the latter representing a ``threshold" halo virial velocity below which most ISM wind mass will leave the halo. We assume $\alpha_{\rm eject}=6$ and $V_{\rm eject}=110$ km s$^{-1}$ following \citet[][and more recent Santa Cruz SAM studies]{somerville08}. Hence, $f_{\rm eject}\times\dot{M}_{\rm ISM,out}$ gives the mass addition rate for the ejected reservoir and the remainder (1-$f_{\rm eject})\times\dot{M}_{\rm ISM,out}$ is deposited into the CGM. The ejected gas can re-accrete into the halo on a Hubble time and become re-eligible for cooling as described earlier.

\section{Analysis}\label{sec:analysis}
Here we describe how we analyze the hydrodynamical simulations and generate semi-analytic predictions for comparison.

\subsection{Generating merger trees and SAM predictions}
We run the Rockstar halo finder \citep{behroozi13a} to generate halo catalogs at each snapshot for the full hydrodynamical FIRE-2 simulations which include both baryonic and DM particles. But since Rockstar will only use DM particles to define virial overdensities and hence halo boundaries, we enable its option to up-weight the DM density field. We adopt the \citet{bryan98} definition of halo virial mass and radius. We only output properties of halos that have at least 100 DM particles associated with them (i.e., within their virial radius); this is the default threshold below which Rockstar discards halos as noise. Next, we run the companion consistent-trees code \citep{behroozi13b} to generate gravitationally-consistent merger trees. This code corrects inconsistencies in the default Rockstar-based merger trees by: (1) removing spurious detections of halos, (2) inserting ``phantom" halos at snapshots where a descendant halo is not identified but should obviously exist due to re-appearance of the halo in a subsequent snapshot, and (3) slightly modifying the positions and velocities of halo centers by comparing to the expected evolution between snapshots based on simple gravitational force calculations. In the end, our halo virial masses and radii agree with those reported in Table 1 of \citet{hopkins18} to within 0.1 dex. 

With the halo merger trees in hand, we run the \citet{somerville15} version of the Santa Cruz SAM with the same observational calibration as used in \citet[][with AGN feedback disabled; see our \autoref{sec:scsam} above for details]{popping19}. Since the SAM has its own model for generating subhalos and predicting their evolution \citep[section 2.1 of][]{somerville08}, we have discarded all subhalos from the merger trees. This is appropriate for our study since we are only focusing on the evolution of the central halo in each of the FIRE-2 zoom simulations (along the most massive progenitor branch) and comparing subhalo modeling is deferred to future work. Note that we are running the SAM on merger trees extracted from the full hydrodynamical simulations, whereas it would be more faithful to use merger trees extracted from corresponding DM-only simulations. However, such DM-only simulations only exist for a subset of the FIRE-2 suite and hence we use the full hydrodynamical suite to increase our sample size (13 halos). In Appendix \ref{sec:dmonly}, we show that none of our conclusions change when we use only the limited DM-only simulation suite. Nevertheless, we emphasize that the only input for the SAM is the dark matter halo merger trees: the SAM is not provided any information about the baryonic properties of the halos.

\subsection{Computing bulk and flow quantities in the simulations}
\begin{figure*}[!htbp]
\begin{center}
\includegraphics[width=0.95\hsize]{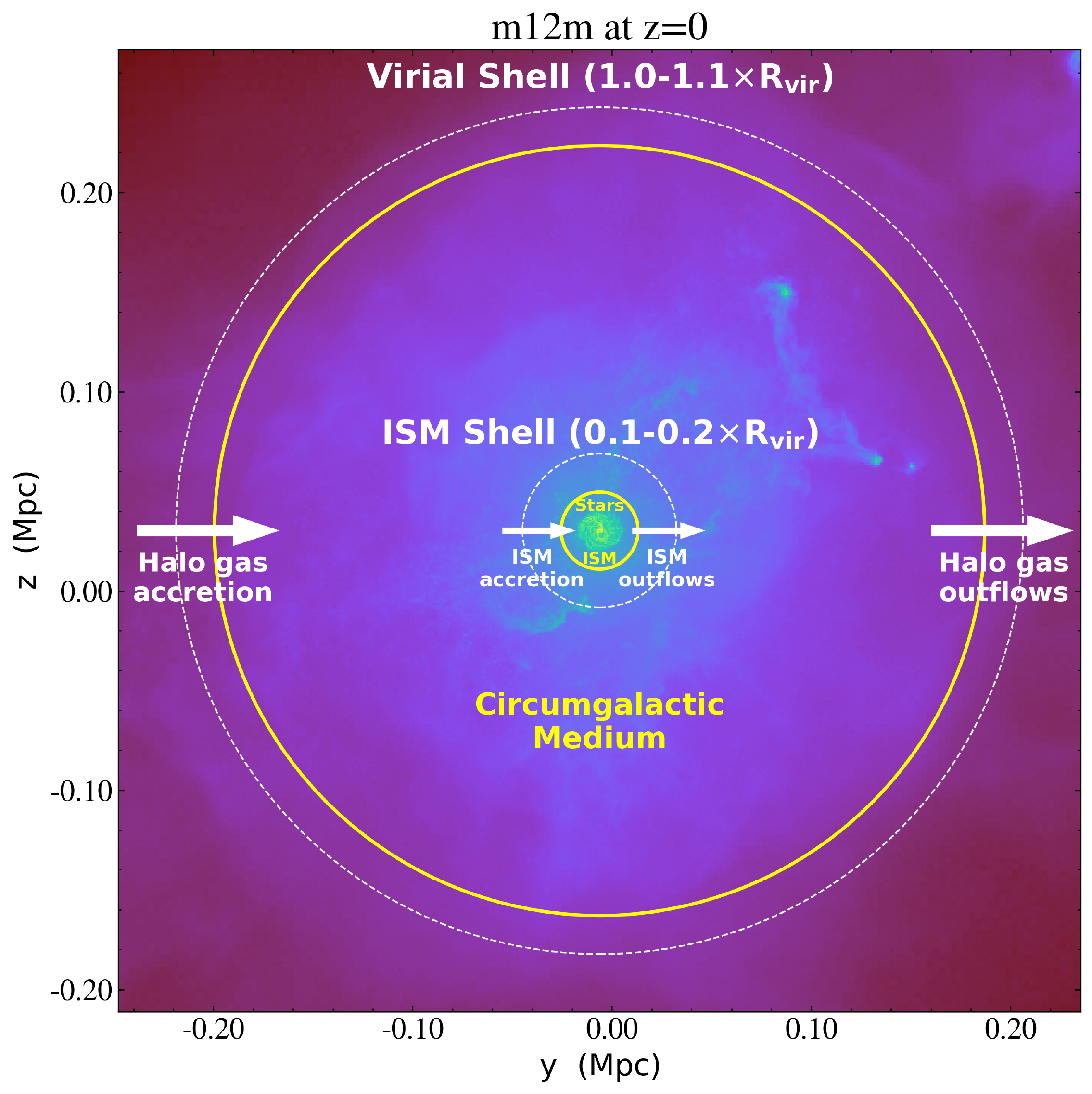}
\end{center}
\caption{An illustration of our zone definitions for analyzing bulk and flow quantities in the FIRE-2 simulations. The background image shows the projected gas density distribution of the MW-mass halo m12m at $z=0$ (purple is low density, with green and yellow representing progressively higher densities, respectively). The solid yellow circles represent the virial radius (outer circle) and the ``edge" of the ISM (inner circle). The dashed white circles demarcate the outer limits of the virial and ISM shells ($1.0-1.1\times R_{\rm vir}$ and $0.1-0.2\times R_{\rm vir}$, respectively) through which mass inflow and outflow rates are computed. Stellar and ISM masses are computed using all star and gas particles within $0.1\times R_{\rm vir}$, respectively, whereas CGM masses are computed using all gas particles between $0.1-1.0\times R_{\rm vir}$. These definitions are well-matched for comparison to the Santa Cruz SAM.}
\label{fig:shells}
\end{figure*}

Our merger trees tell us the center position and radius of the central halo in every snapshot, as well as many other halo properties. With this information, we can use the simulation particle data to compute the baryonic properties of the central halo along the most massive progenitor branch. In \autoref{fig:shells}, we illustrate how we compute bulk masses and differential mass flow quantities in different ``zones" for the hydrodynamical data. The definitions of these zones are well-matched to the SAM for comparison. We define the stellar mass as the sum of the masses of all star particles within $0.1R_{\rm vir}$. We also define the ISM mass as the sum of all gas particle masses within $0.1R_{\rm vir}$. The CGM mass is defined as the sum of all gas particle masses between $0.1R_{\rm vir}$ and $1.0R_{\rm vir}$, irrespective of temperature, density, etc. We already have the dark matter halo mass from the merger trees, which is based on the sum of all DM particle masses within $1.0R_{\rm vir}$. These constitute our main integrated mass measurements. We also compute instantaneous global galaxy SFRs by summing up the predicted instantaneous SFRs of all individual gas particles within $0.1R_{\rm vir}$. We have also computed time-averaged SFRs based on adding up stars with ages younger than 20 Myr, 100 Myr, and 1 Gyr, and find good agreement with the instantaneous gas-based measurements after boxcar smoothing. By default, we use the instantaneous gas-based measurements since these are closer in definition to what the Santa Cruz SAM predicts.

We adopt the approach of \citet{muratov15} to measure instantaneous mass flow rates within radial shells. Specifically, for all particles within a given radial shell, we compute their radial velocities including the contribution from the Hubble flow (this is generally minor but it can have a differentially larger effect in halo outskirts). We define all particles with negative halo-centric radial velocities as inflowing, and similarly all particles with positive radial velocities as outflowing. Then, the mass inflow rate for a given radial shell is the weighted sum of the individual particle mass fluxes using only the particles with negative radial velocities: 
\begin{equation}
\dot{M} = \sum_i \frac{m_i |v_{r,i}|}{dL}\;.
\end{equation}
Here, $m_i$ is the mass of particle $i$ in the shell, $|v_{r,i}|$ is the absolute value of its radial velocity and $dL$ is the shell width. An analogous calculation is done separately for the mass outflow rate using only particles in the shell with positive radial velocities. In this way, particles with slower velocities contribute less mass flux than those with higher velocities (for a given particle mass), and the dependence of the mass flux measurement on the shell width is accounted for. 

We make mass inflow and outflow rate measurements in two spherical shells at every snapshot. We define a ``virial shell" that extends from $1.0-1.1R_{\rm vir}$ and an ``ISM shell" that extends from $0.1-0.2R_{\rm vir}$. The widths of both shells are thus $0.1R_{\rm vir}$. We have carried out extensive convergence tests for the location and width of each shell. In short, the definition of the virial shell is robust to reasonable changes in the centering and width, especially since we take the halo virial radius as a given from the merger tree. On the other hand, the definition of the ISM shell is more arbitrary since there is no obvious ISM ``edge" in either the simulation or the SAM. The ISM shell width represents a good compromise between mitigating Poisson noise, systematically missing the fastest moving particles, and accurately capturing the bulk flow of mass as a function of radius across snapshots. The ISM shell is located at a considerable distance from the ISM which means that there can be contamination from ambient inner CGM material or fountain flows. However, if the shell is placed too close to the ISM, then the flow measurements can also be contaminated by the dense ISM. Without imposing more sophisticated criteria to select truly escaping or accreting ISM particles and accounting for the complicated geometrical evolution of galaxies, there thus needs to be a compromise. Overall, we find that our choice of shell definitions are sensible for comparison to the SAM and for measuring flow rates out to $z\sim10$ \citep[and they are also standard in the literature;][]{fauchergiguere11,muratov15}. 

Another way to derive mass flow rates is via particle tracking, which also has the advantage of providing information about recycling distances and timescales. \citet{anglesalcazar17}, \citet{hafen19} and \citet{hafen20} have already performed this particle tracking analysis for both the FIRE-1 and FIRE-2 simulations, and we will discuss their results in the context of our work. Note that throughout this paper we will use the ``pure" inflow and outflow rates separately instead of the net inflow rate (i.e., inflow minus outflow). We do not attempt to excise satellites whose own orbits and outflows can contaminate our flow measurements for the central halo. Leaving satellites in may also bias our computed CGM masses a bit high, although it does make for a more consistent comparison to the SAM (which transfers the CGM of subhalos to that of the parent halo; \autoref{sec:scsam}).

\section{Results}\label{sec:results}
Here we present results from our comparison between the SAM and FIRE-2 predictions. We will first present the bulk mass quantities and then the mass flow quantities to better pinpoint any discrepancies.\footnote{Unless otherwise noted, we boxcar smooth all time series by $\pm1$ Gyr to facilitate visual comparison. While this smoothing does wash out features on much smaller timescales, our conclusions would broadly remain the same had we not smoothed the data.}

\subsection{Stellar, ISM and CGM mass scalings at $z=0$}\label{sec:obs}
\begin{figure*}[!htbp]
\begin{center}
\includegraphics[width=0.99\hsize]{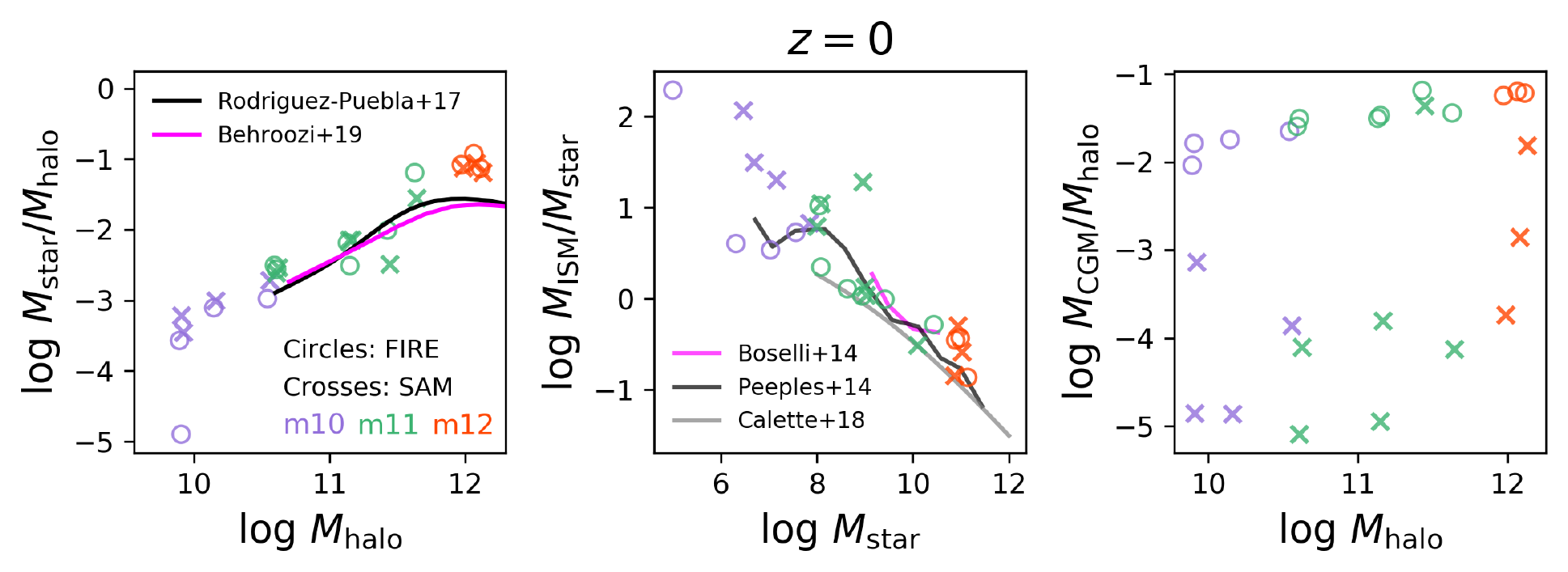}
\end{center}
\caption{Our FIRE-2 measurements (circles) and SAM predictions (crosses) for stellar-to-halo mass ratios (left), ISM-to-stellar mass ratios (middle), and CGM-to-halo mass ratios (right) at $z=0$. Halos are colored according to their mass bin (low-mass dwarfs in purple, intermediate-mass dwarfs in green and MW-mass halos in red). We also show observationally inferred scaling relations for median stellar-to-halo mass ratios \citep{rodriguezpuebla17,behroozi19} and ISM-to-stellar mass ratios \citep{boselli14,peeples14,calette18}. We do not show observational estimates of CGM-to-halo mass ratios since they are highly uncertain and the SAM is not calibrated to match observed CGM properties. The SAM and FIRE-2 agree relatively well with each other and with these observations in terms of stellar-to-halo and ISM-to-stellar mass ratios at a fixed mass \citep[the ISM-to-stellar mass ratios predicted by the SAM for low-mass dwarfs would agree better with observations if we only included the cold atomic and molecular phases;][]{popping14,somerville15}. By comparison, the SAM and FIRE-2 predictions for CGM-to-halo mass ratios disagree dramatically with each other, especially for the dwarfs where the SAM predictions are generally lower by $\sim3$ orders of magnitude.}
\label{fig:obs}
\end{figure*}

We begin by showing mass-dependent scaling relations at $z=0$ for the SAM, FIRE-2 and observations in \autoref{fig:obs}. We focus on stellar-to-halo, ISM-to-stellar and CGM-to-halo mass ratios (as a function of the denominator mass; no boxcar smoothing). We include comparisons to median stellar-to-halo mass relations derived from halo abundance matching \citep[from][]{rodriguezpuebla17,behroozi19} and ISM-to-stellar mass relations from observations \citep[from][]{boselli14,peeples14,calette18}. We do not include observationally inferred CGM-to-halo mass ratios because there are large systematic uncertainties associated with measuring the total CGM masses of galaxies. Versions of the stellar-to-halo and (atomic plus molecular) ISM-to-stellar mass ratio relations are used to calibrate the SAM.\footnote{As mentioned in \autoref{sec:scsam}, we have disabled AGN feedback in the SAM for the sake of a fair comparison with FIRE-2, even though the SAM relies on AGN feedback to agree with observations for MW and higher mass halos (there are no appreciable effects for dwarfs in the SAM). We find that enabling AGN feedback in the SAM decreases the stellar and ISM masses of MW halos by a couple tenths of a dex, and increases their CGM masses by more than a dex \citep[owing to quasar winds ejecting ISM mass and radio jet heating offsetting CGM cooling;][]{somerville08}. As expected, this brings the SAM MW halos into even better qualitative agreement with the observations shown.} We emphasize that these observational comparisons are purely illustrative: we have not made an effort to properly generate mock observables and there are a few caveats. First, the stellar-to-halo mass relations based on subhalo abundance matching are only valid at $M_{\rm vir}\gtrsim10^{10.5}M_{\odot}$ due to the resolution of the DM simulations used so we cannot comment on the low-mass dwarfs \citep[but see][]{wheeler19}. In addition, we do not make any cuts on ISM gas phase for the SAM and FIRE-2 predictions even though the observationally inferred ISM-to-stellar mass ratios plotted in \autoref{fig:obs} account for only the cold atomic and molecular gas phases (i.e., HI and H$_2$). This is done to prevent confusion throughout the rest of this paper where we will simply want to compare the total ISM masses between the SAM and FIRE-2 (neglecting the physics of multi-phase gas partitioning, which is beyond the scope of this paper). Note, however, that the SAM predictions for the cold atomic and molecular ISM gas masses alone (excluding HII) have been shown agree well with observations at $z\sim0$ \citep{popping14,somerville15}.

Overall we find that the SAM and FIRE-2 predictions agree relatively well with each other and with observations for the stellar-to-halo and ISM-to-stellar mass ratios at fixed mass, but disagree dramatically on CGM-to-halo mass ratios. In detail, the stellar-to-halo mass ratios generally agree with the abundance matching relations within a factor of two for both the SAM and FIRE-2.\footnote{Figure 7 of \citet[][]{hopkins18} shows even better agreement for the m12 halos. Our virial and stellar masses agree with those of \citet[][Table 1]{hopkins18} within 0.1 dex, but our stellar masses are slightly larger whereas our virial masses are slightly smaller. Hence, our estimate of the stellar-to-halo mass ratio itself will be biased higher than theirs. The virial mass disagreement can likely be attributed to our different halo finders whereas the stellar mass difference is likely due to our different assumed integration radius. We use $0.1R_{\rm vir}$ for simplicity but they use a more refined, slightly smaller definition (three times the iteratively computed 3D stellar half-mass radius).} We do not attempt to extend the abundance matching relations to low-mass dwarfs. As for the ISM-to-stellar mass ratio, the SAM and FIRE-2 agree relatively well with each other and with the observations for the m11 and m12 halos. This is remarkable since no attempt was made to force the SAM to reproduce FIRE-2, and FIRE-2 itself was not calibrated to match observations. However, for the m10 halos, the SAM is higher than FIRE-2 by up to a factor of ten. This order of magnitude disagreement persists if we separately compare just the cold ISM mass (atomic plus molecular; defined crudely in FIRE-2 as all gas particles at $<0.1R_{\rm vir}$ with $T<10^4$ K) or the warm ionized gas mass (HII; defined crudely in FIRE-2 using gas particles at $<0.1R_{\rm vir}$ that have $T=10^4-10^5$ K).

Strikingly, all of these differences are eclipsed by discrepancies in the CGM-to-halo mass ratios: the SAM predictions are orders of magnitude lower than FIRE-2, with the deficit being systematically larger for lower mass halos ($\sim3$ orders of magnitude). The ability to agree relatively well on stellar and ISM mass but disagree by orders of magnitude on CGM mass reflects the flexibility allowed in phenomenological models for the baryon cycle. In our case, this flexibility arises because the SAM is not calibrated to match the observed CGM masses of galaxies (which are highly uncertain; it is not clear whether the bulk of extragalactic, non-ISM gas bound to halos is located within or outside of those halos).

\subsection{Stellar mass histories}
\begin{figure*}[!htbp] 
\begin{center}
\includegraphics[width=0.9\hsize]{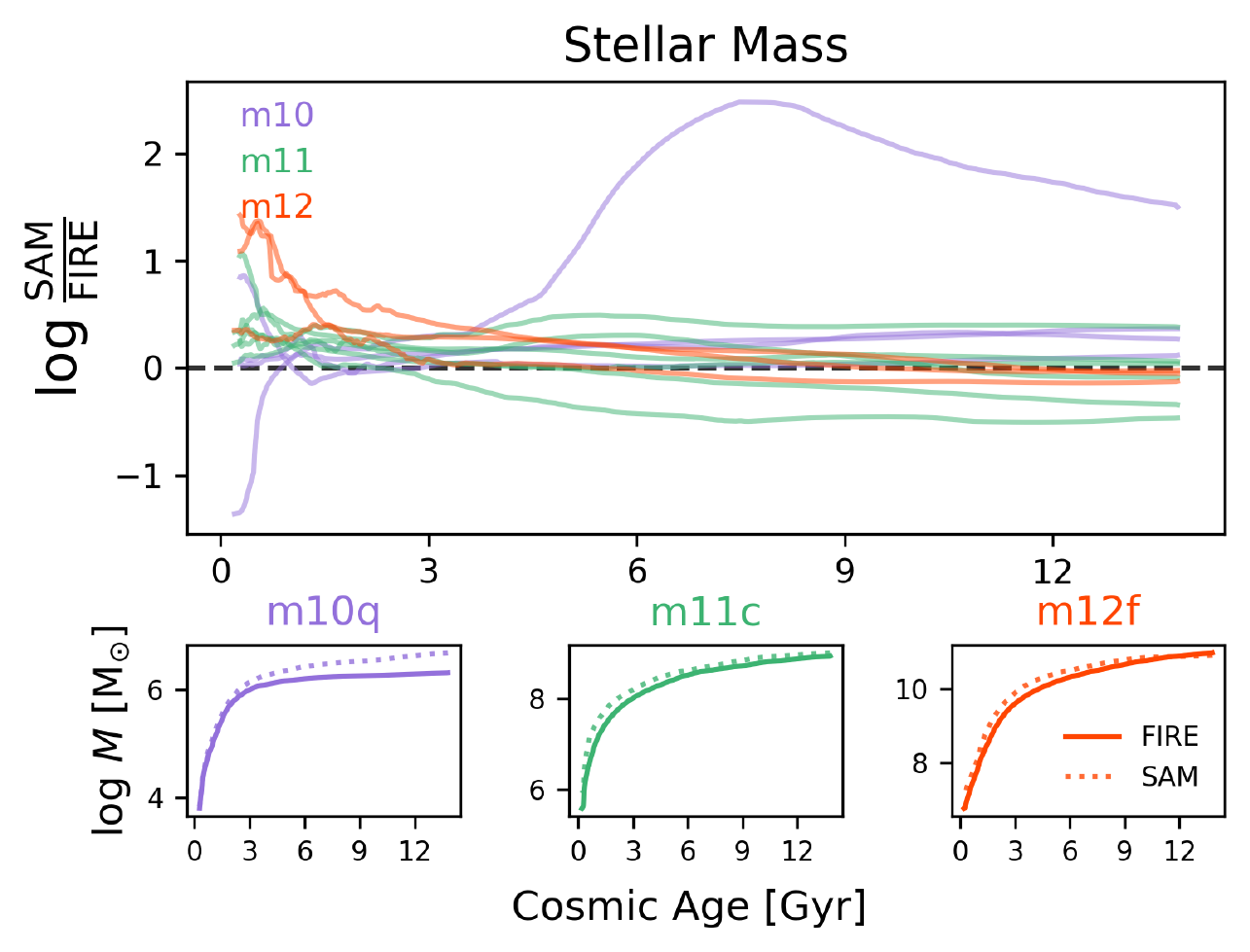}
\end{center}
\caption{Comparison of the stellar mass assembly history measured in FIRE-2 and as predicted by the SAM. Top: logarithmic ratio of the SAM and FIRE-2 time series color-coded by mass bin (m10 halos in purple, m11 halos in green, and m12 halos in red). Bottom: individual stellar mass assembly histories for one representative halo from each mass bin (m10q left, m11c middle, m12f right).  All time series are smoothed over $\sim1$ Gyr for easier visual comparison. With one exception (m10v), the SAM generally reproduces the FIRE-2 stellar mass assembly histories within a factor of two.}
\label{fig:mass_stars}
\end{figure*}

While the previous comparison of mass-dependent scaling relations at $z=0$ is already suggestive of significant model discrepancies, it is insightful to compare the full time evolution of various properties. We start with stellar mass in \autoref{fig:mass_stars}. Overall, the SAM and FIRE-2 agree roughly within a factor of two. Although the SAM was tuned to reproduce the $z=0$ stellar mass function, it is \emph{not} tuned to reproduce observations at earlier cosmic epochs, although its predictions have been shown to be in reasonably good agreement with observations such as luminosity and stellar mass functions out to $z\sim10$ \citep{somerville15,yung19a,yung19b}. Two trends are evident: the SAM tends to predict higher stellar masses than FIRE-2 at early times in MW-mass halos (by up to a factor of 10) and to a lesser extent in the m11 halos; and it also predicts higher stellar masses than FIRE-2 in the low-mass dwarfs at late times (but by less than a factor of two, except for the remarkably late-forming halo m10v, which we will discuss later).

\subsection{Star formation stochasticity} 
\begin{figure}[!htbp]
\begin{center}
\includegraphics[width=0.99\hsize]{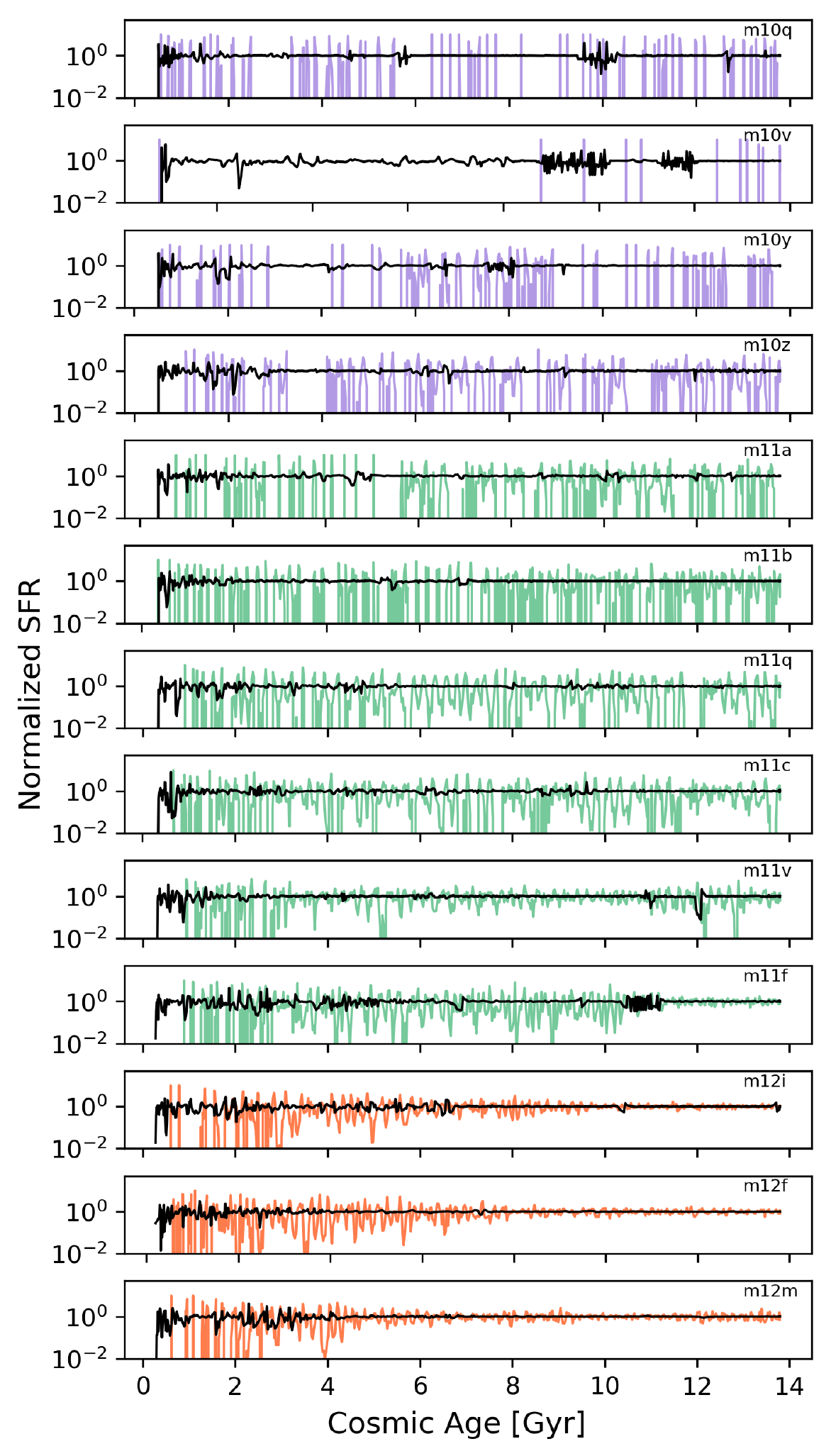}
\end{center}
\caption{Normalized SFHs for FIRE-2 measurements (colored curves) and SAM predictions (black). The time series are the instantaneous SFHs divided by the corresponding 100 Myr boxcar-smoothed SFHs. The halos are ordered based on increasing $z=0$ halo mass from top to bottom (m10 halos in purple, m11 in green, m12 in red). The m10 and m11 FIRE-2 halos are bursty at all times, and the m12 halos are bursty at early times but not late times. In contrast, the SAM predicts much less SFH stochasticity.}
\label{fig:mdot_sfr}
\end{figure}

That the overall stellar mass assembly histories agree already suggests that the star formation histories (SFHs) must also agree when averaged over sufficiently long timescales. Indeed, we find that this is generally the case. However, on shorter timescales ($\sim100$ Myr), the behavior of the SAM and FIRE-2 SFHs are very different. In \autoref{fig:mdot_sfr}, we show the normalized SFHs of all 13 FIRE-2 halos and include the SAM predictions. As already shown by \citet{sparre17} and \citet{fauchergiguere18}, the FIRE-2 m10 and m11 halos have bursty SFHs at all times, whereas the more massive m12 halos are only bursty at early times ($z\gtrsim1$ corresponding to cosmic ages $\lesssim6$ Gyr) and settle into a more steady state at later times \citep[see also][]{muratov15,anglesalcazar17,ma17}. These trends are not predicted by the SAM, in which there is systematically much lower SFH variability compared to FIRE-2.

\subsection{ISM mass histories}
\begin{figure*}[!htbp]
\begin{center}
\includegraphics[width=0.9\hsize]{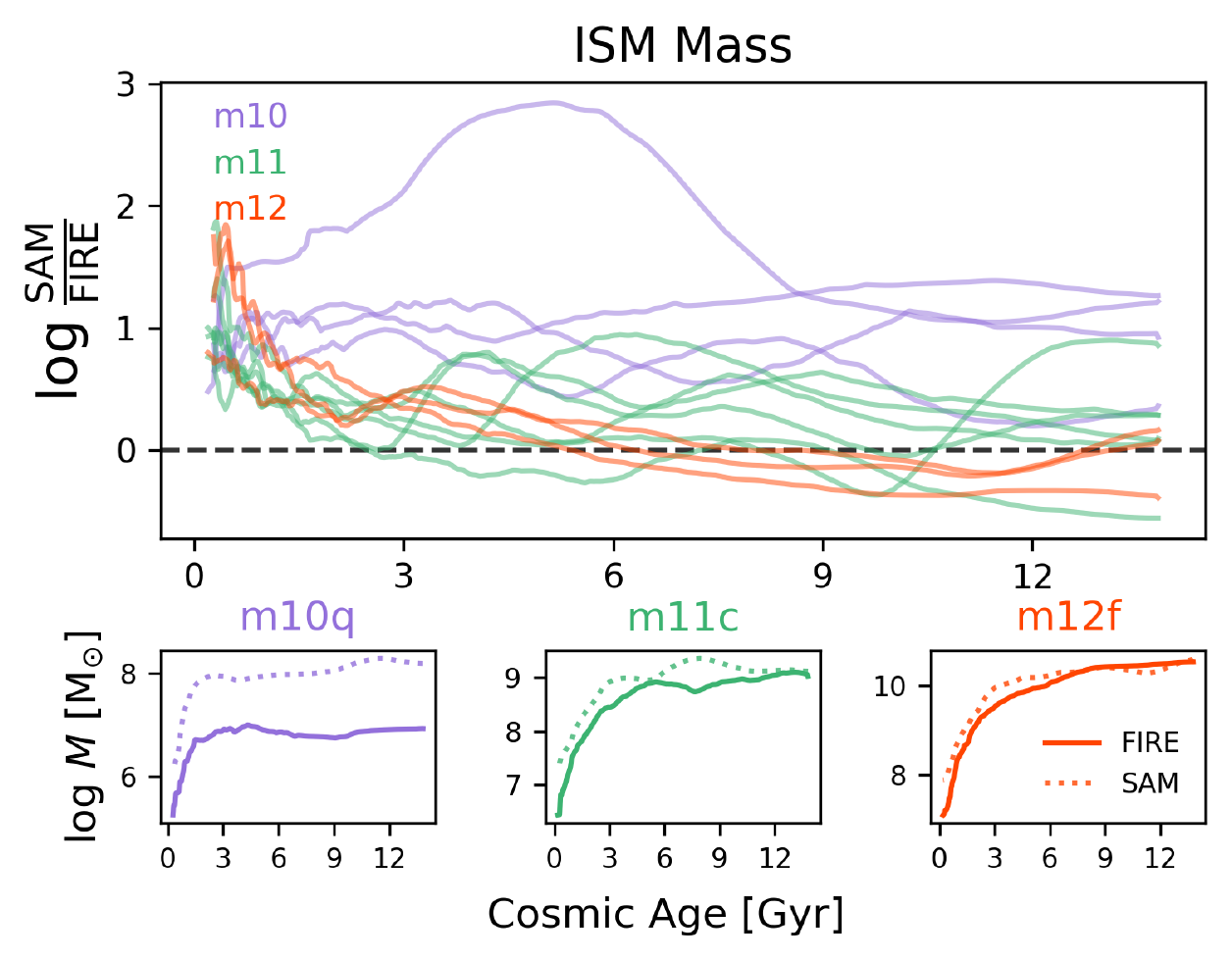}
\end{center}
\caption{Similar to \autoref{fig:mass_stars} but now for ISM mass as a function of cosmic age. The SAM agrees relatively well with FIRE-2 for the m12 halos except at very early times, and this is also true for the m11 halos, albeit with more scatter. But the systematic discrepancy for the m10 halos remains at the order of magnitude level at all times, consistent with the $z=0$ SAM excess in \autoref{fig:obs}.}
\label{fig:mass_ism}
\end{figure*}

\autoref{fig:mass_ism} now compares the ISM mass histories between FIRE-2 and the SAM.
Overall we see more disagreement here. The SAM predicts higher ISM masses than FIRE-2 in halos of all masses at very early times (up to a factor of ten). The SAM ISM masses are higher by at least a factor of $\sim 5$-10 in nearly all the m10 halos over all of cosmic time (as discussed in \autoref{sec:obs}, these differences persist if we only consider the cold or warm ionized components). The m11 ISM masses predicted by the SAM tend to be higher than FIRE-2 by about a factor of 2-3 over most of cosmic time. The MW mass halos mostly show good agreement between the two methods (within a factor of 2) after a cosmic age of about 6 Gyr.

\subsection{CGM mass histories}
\begin{figure*}[!htbp] 
\begin{center}
\includegraphics[width=0.9\hsize]{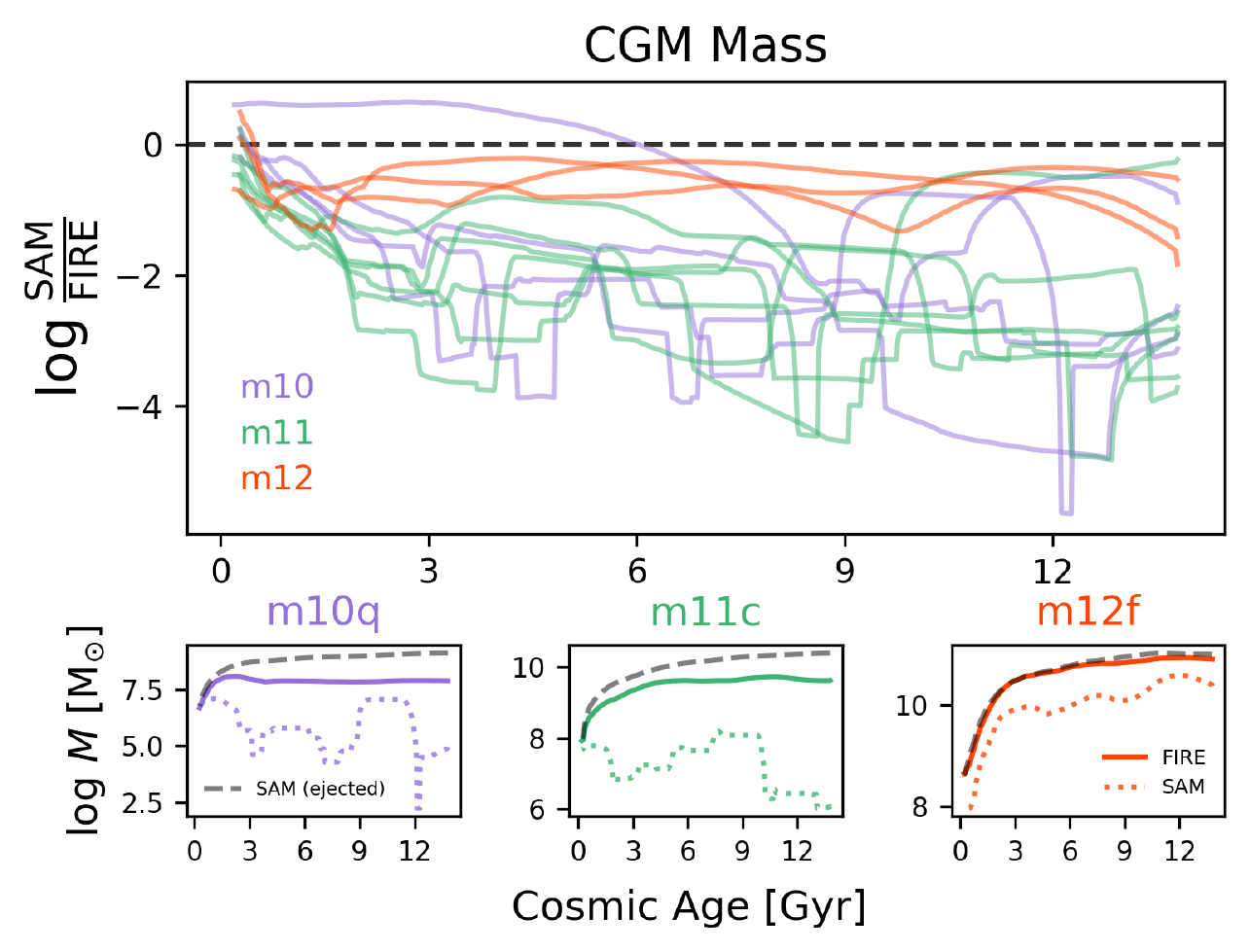}
\end{center}
\caption{Similar to \autoref{fig:mass_stars} but now for CGM mass. The SAM predicts much lower CGM masses than FIRE-2 for all halos, with the deficit being worse for the m10 and m11 halos ($\sim3-4$ orders of magnitude). In the individual example halo panels (bottom), we overplot the mass of the ``ejected gas" reservoir (dashed gray lines) and see that it alone is larger than the FIRE-2 CGM mass. The SAM CGM masses are likely very low because most of the gas resides in this ejected reservoir (i.e., the SAM predicts that most of the extragalactic yet bound gas resides outside of the halo).}
\label{fig:mass_cgm}
\end{figure*}

Next we will compare the ``CGM" mass predicted by the SAM and as measured in FIRE-2. In \autoref{fig:mass_cgm}, we plot the time evolution of the CGM mass in FIRE-2 and in the SAM. It is immediately obvious that the CGM mass is much lower in the SAM than in FIRE-2 in all halos at all times. The CGM mass is $\sim3-4$ orders of magnitude lower in the SAM than in FIRE-2 for the m10 and m11 halos. While the discrepancy is smaller for the m12s, the SAM still has lower CGM masses than FIRE-2 by $\sim1$ order of magnitude. The ``boxy" trajectories for CGM mass in the dwarfs are likely an artifact of the SAM CGM cooling model (the CGM mass may be constant when $R_{\rm cool}>R_{\rm vir}$ and the halo gas inflow rate equals the ISM inflow rate, assuming outflows and subhalo accretion are a negligible source of CGM mass growth; \autoref{sec:coolingmodel}). 

For context, we also plot the time evolution of the ``ejected" gas mass reservoir for the individual example SAM halos, and see that it dominates over the CGM mass. Most of this extragalactic (i.e., non-ISM but still bound) gas resides outside of the halo in the SAM, and its mass alone agrees better with the FIRE-2 CGM mass (especially for the MW halos). 

\subsection{Halo baryon fraction evolution}
\begin{figure*}[!htbp] 
\begin{center}
\includegraphics[width=0.9\hsize]{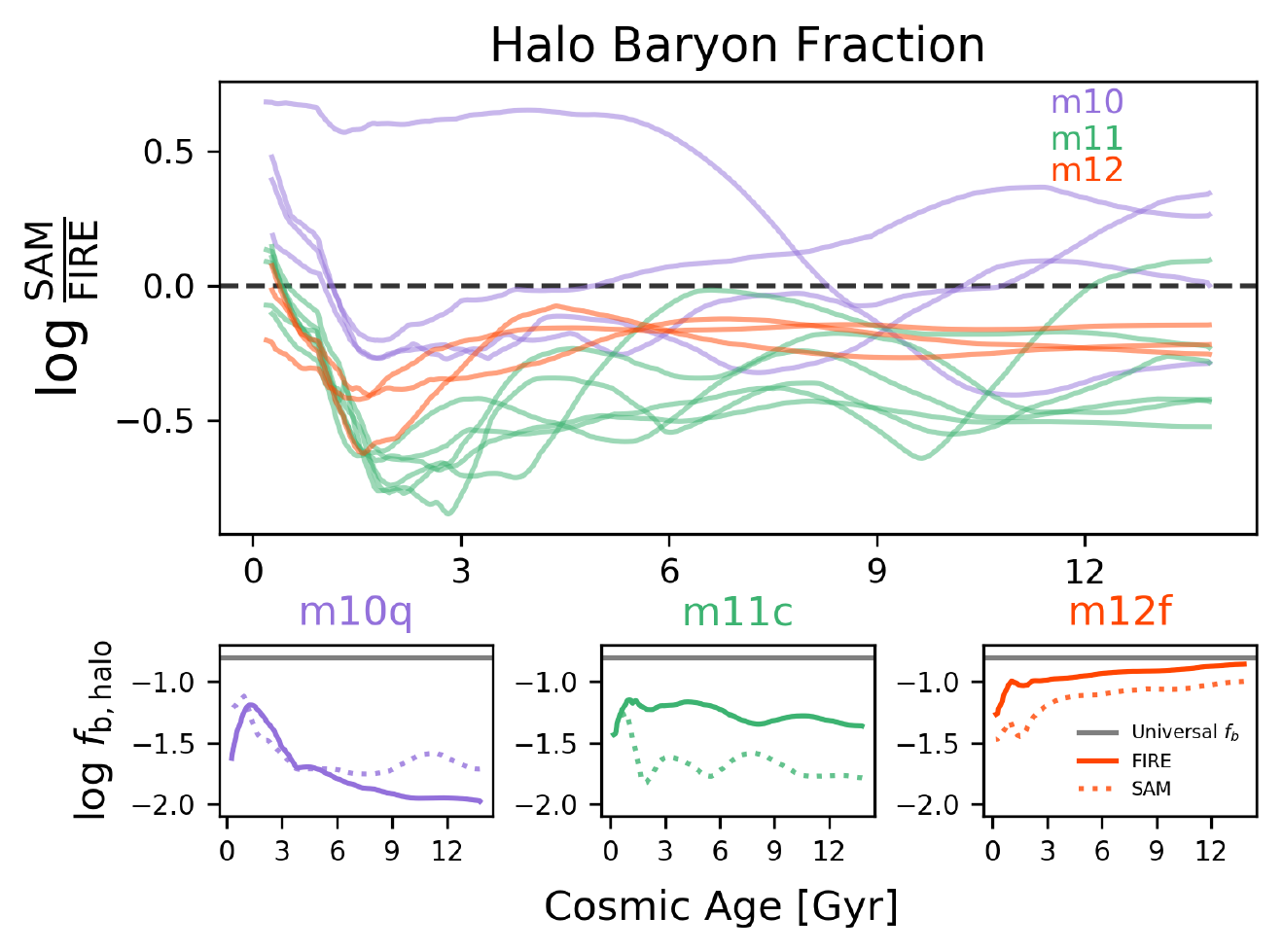}
\end{center}
\caption{Similar to \autoref{fig:mass_stars} but now for the bulk halo baryon fractions. The SAM reproduces the general trend in FIRE-2: lower mass halos are more depleted of baryons than higher mass halos, relative to the universal baryon fraction \citep[horizontal gray lines in the bottom panels; $f_{\rm b}=0.158$ according to][]{planck16}. There is relatively good agreement between the SAM and FIRE-2, with differences at the factor of $\sim2-3$ level at most (despite order of magnitude differences in ISM and CGM mass; note the much smaller $y$-axis range in this figure compared to \autoref{fig:mass_ism} and \autoref{fig:mass_cgm}).} 
\label{fig:fb_evolution}
\end{figure*}

Finally, it is useful to combine the three previous bulk mass quantities and define the bulk halo baryon fraction: 
\begin{equation}
f_{\rm b,halo}=\frac{M_{\rm stars}+M_{\rm ISM}+M_{\rm CGM}}{M_{\rm stars}+M_{\rm ISM}+M_{\rm CGM}+M_{\rm DM}}\;. 
\end{equation}
Consistent with \citet{muratov15}, \citet{fitts17} and \citet{hafen19}, in \autoref{fig:fb_evolution} we show that lower mass FIRE-2 halos are more depleted of baryons than higher mass halos, relative to the universal baryon fraction \citep[$f_{\rm b}=0.158$ according to][]{planck16}. The SAM reproduces this overall trend. In more detail, the SAM predictions relative to FIRE-2 are systematically lower for the m11 and m12 halos and similar or higher for the m10 halos. However, the differences are roughly at the factor of $\sim2-3$ level at most and primarily driven by the CGM mass deficit in the SAM (which predicts that most of the extragalactic/non-ISM bound gas resides outside of the halos rather than in the CGM). The main reason why the m10 halos tend to have somewhat similar (or higher at late times) baryon fractions in the SAM than FIRE-2 is because their CGM mass deficit is somewhat offset by their ISM mass excess. It is interesting that any order of magnitude discrepancies in the individual mass components (namely ISM and CGM mass) manifest as relatively inconsequential differences for the halo baryon fraction, suggesting that this is an ambiguous quantity to interpret on its own.

\subsection{Halo mass inflow rates}
\begin{figure*}[!htbp] 
\begin{center}
\includegraphics[width=0.9\hsize]{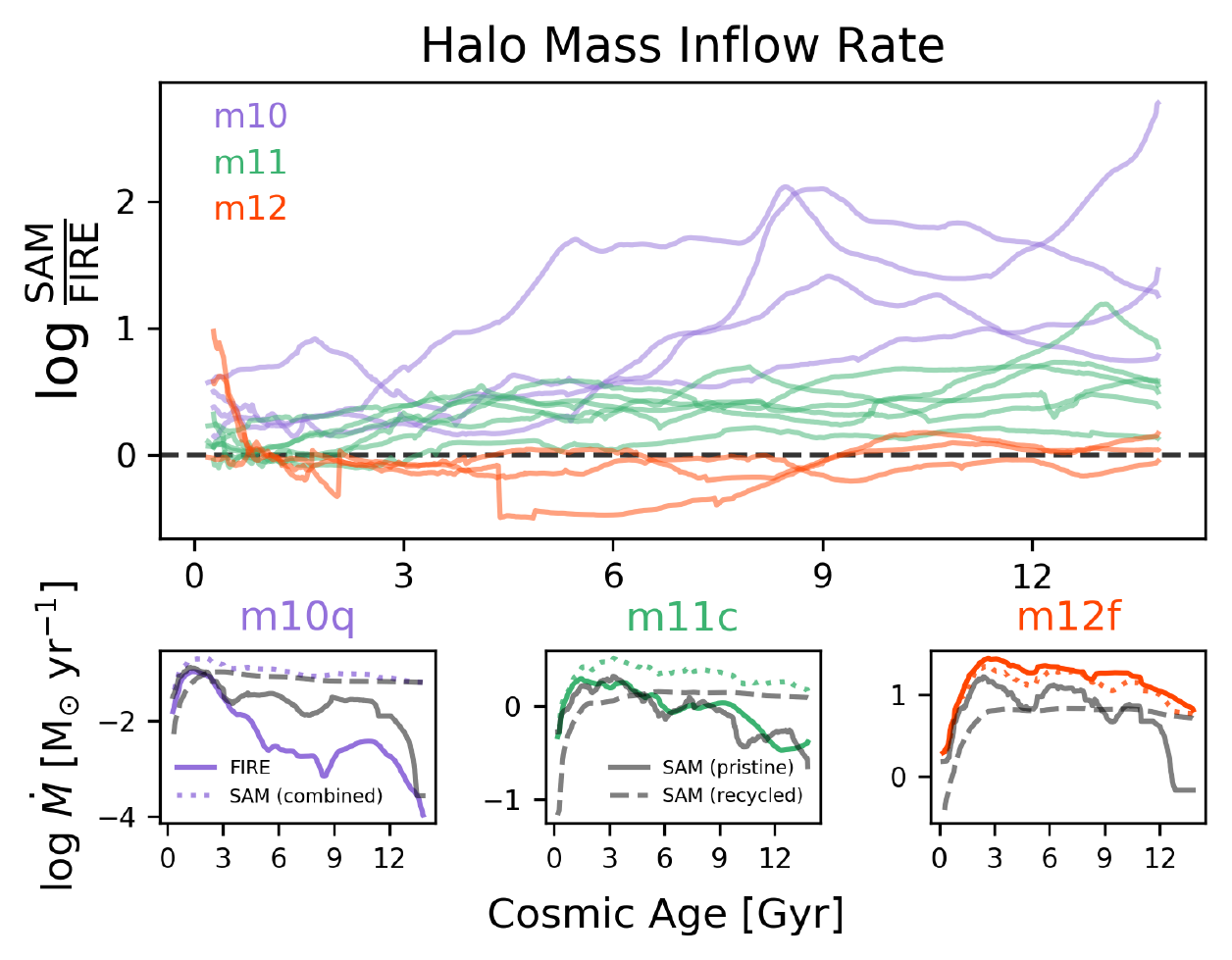}
\end{center}
\caption{Similar to \autoref{fig:mass_stars} but now for halo gas accretion rate. In the bottom panels, we also plot the SAM halo gas accretion rate split into pristine accretion (solid gray) and re-accretion of previously ejected gas (dashed gray). The SAM matches the MW-mass halo gas accretion rates relatively well, but predicts significantly higher values for the dwarfs (by $\sim1-2$ orders of magnitude). This excess accretion in the SAM is primarily driven by its high ejected gas re-accretion rate, but the pristine accretion by itself is still higher than FIRE-2 for the low-mass dwarfs (m10q is representative).}
\label{fig:mdot_in_halo}
\end{figure*}

In order to better pinpoint what is driving the trends in the bulk quantities above, we now turn to a comparison of differential quantities, namely the corresponding mass inflow and outflow rates for the ISM and CGM. We begin with the halo mass inflow rate in \autoref{fig:mdot_in_halo}. For the MW-mass halos, the SAM agrees relatively well with FIRE-2 effectively at all times. But for progressively lower mass halos, the SAM predicts systematically higher halo gas accretion rates than measured in FIRE-2, with the discrepancy getting somewhat worse with time. For the m11 halos, the SAM is higher than FIRE-2 by more than a factor of two, and for the m10 halos, the SAM is higher than FIRE-2 by more than a factor of ten. 

We can gain further insight by splitting the SAM halo gas accretion into pristine accretion versus re-accretion of gas that was previously ejected from the halo due to stellar feedback. It then becomes obvious that the re-accretion rate dominates over the pristine accretion rate in the dwarfs (see gray lines in the bottom panels of \autoref{fig:mdot_in_halo}; these halos are representative). Hence, the trend that the overall halo mass inflow rate is higher in the SAM than FIRE-2 for dwarfs is primarily driven by the high re-accretion rates. However, the pristine SAM accretion rate itself can still be significantly higher than FIRE-2 for the low-mass dwarfs, which may reflect preventative feedback not modeled by the SAM. Finally, for the MW-mass halos, the pristine accretion generally dominates over re-accretion, which is sensible since most stellar-driven winds cannot escape the potential well of these more massive halos (\autoref{eqn:feject}). However, there can be dips in the pristine accretion that reflect the underlying DM halo merger history.\footnote{For a central halo that experiences a merger, the halo mass will generally show a sharp jump because the halo finder suddenly assigns to the central halo all the particles belonging to the recently accreted subhalo. The subsequent DM accretion can be lower by comparison, especially while the halos have not fully coalesced.} Coincidentally, these dips are generally compensated for by the re-accretion rate, leading to overall agreement with the FIRE-2 halo inflow rates for the MW-mass halos (as seen for the example m12f halo).

\subsection{ISM inflow rates}
\begin{figure*}[!htbp] 
\begin{center}
\includegraphics[width=0.9\hsize]{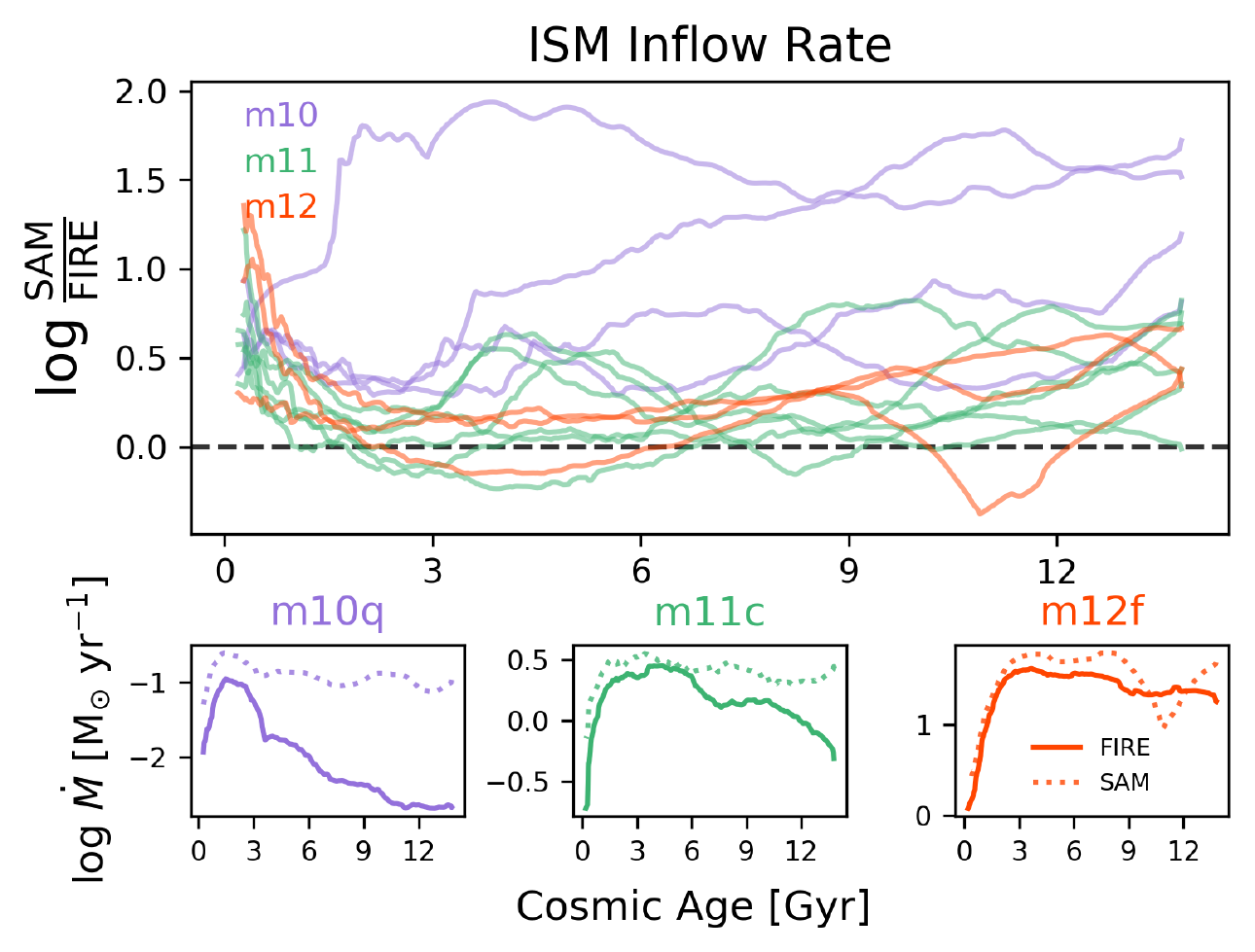}
\end{center}
\caption{Similar to \autoref{fig:mass_stars} but now for ISM gas accretion rate. The SAM generally has higher ISM gas accretion rates than measured in FIRE-2, and this discrepancy is preferentially worse for the lower mass halos (up to two orders of magnitude). Even the m12 halos at late times have about a factor of two higher ISM accretion rates in the SAM than in FIRE-2.}
\label{fig:mdot_in_ism}
\end{figure*}

\autoref{fig:mdot_in_ism} compares the ISM accretion rate between the SAM and FIRE-2. The SAM predicts much higher ISM accretion rates compared to FIRE-2. For the m10 halos, the SAM is higher than FIRE-2 by more than a factor of 10 whereas for the m11 and m12 halos the SAM is larger than FIRE-2 by more than a factor of two (especially at late times).

\subsection{ISM outflow rates}
\begin{figure*}[!htbp] 
\begin{center}
\includegraphics[width=0.9\hsize]{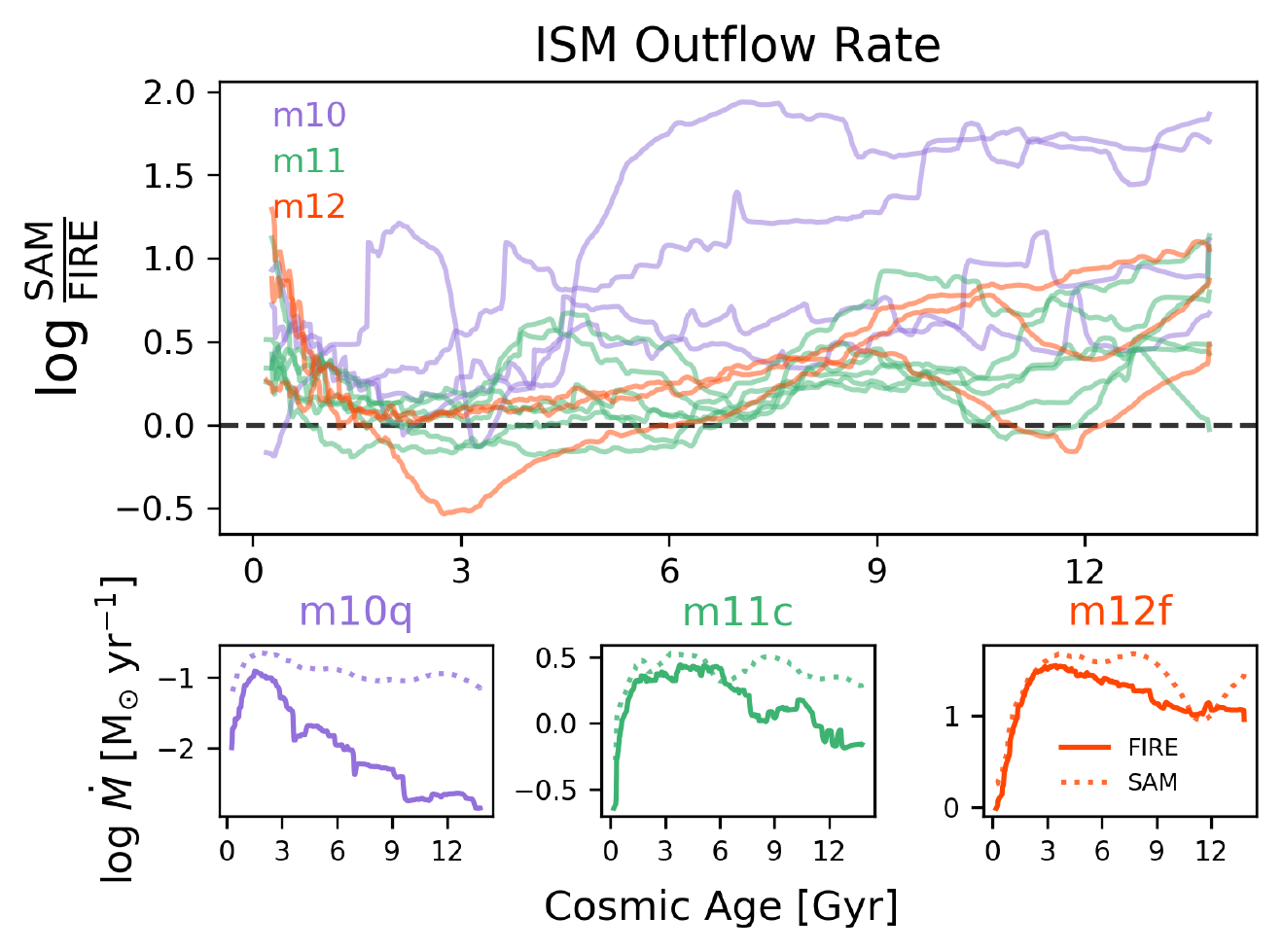}
\end{center}
\caption{Similar to \autoref{fig:mass_stars} but now for the ISM mass outflow rate. The SAM has higher ISM outflow rates than FIRE-2, with the discrepancy becoming larger than a factor of two for most halos by late times. This is necessary in the SAM to prevent excess star formation and match the observed stellar mass function.}
\label{fig:mdot_out_ism}
\end{figure*}

Next, we turn to the ISM mass outflow rate in \autoref{fig:mdot_out_ism}. The SAM ejects much more gas from the ISM than FIRE-2, with the discrepancy being more than a factor of 2 for most halos at most times. This is expected because if the SAM is to match the SMHM relation at $z=0$, then it must remove the excess accreted ISM gas via more efficient stellar feedback. Indeed, we verified that, on average, the net ISM inflow rates (inflow minus outflow) agree relatively well between the SAM and FIRE-2, with some slight discrepancies for the dwarfs (related to their excess ISM masses in \autoref{fig:mass_ism}). However, the issue is that the SAM and FIRE-2 are achieving their similar net inflow rates in different ways.

\subsection{Halo outflow rates}
\begin{figure*}[!htbp] 
\begin{center}
\includegraphics[width=0.9\hsize]{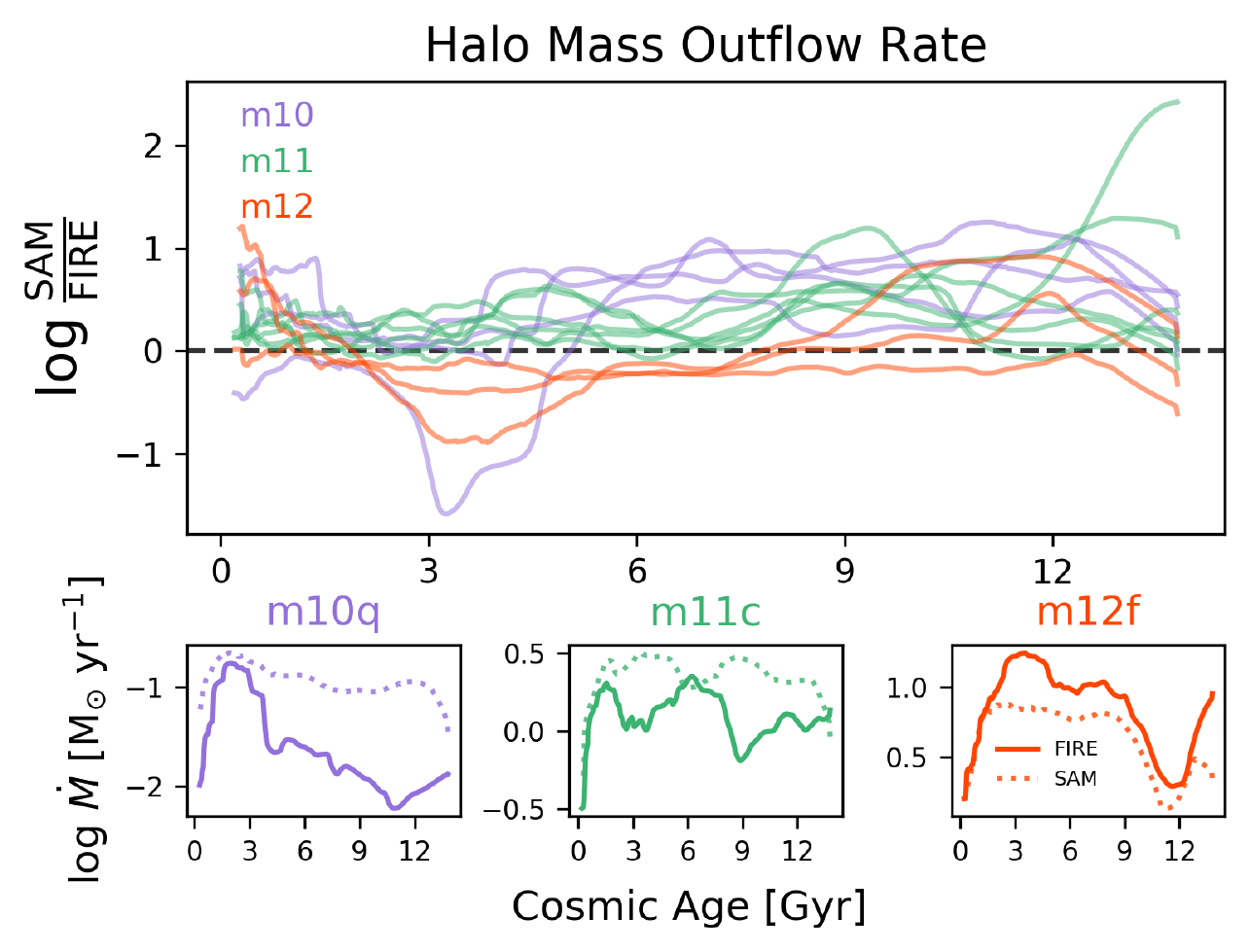}
\end{center}
\caption{Similar to \autoref{fig:mass_stars} but now for the halo mass outflow rates. The SAM has higher halo mass outflow rates than the FIRE-2 measurements. This is not surprising because the halo mass outflow rate in the SAM is a halo circular velocity-dependent re-scaling of the ISM outflow rate, and the latter was already shown to be much higher than FIRE-2.}
\label{fig:mdot_out_halo}
\end{figure*}

\begin{figure*}[!htbp]
\begin{center}
\includegraphics[width=0.9\hsize]{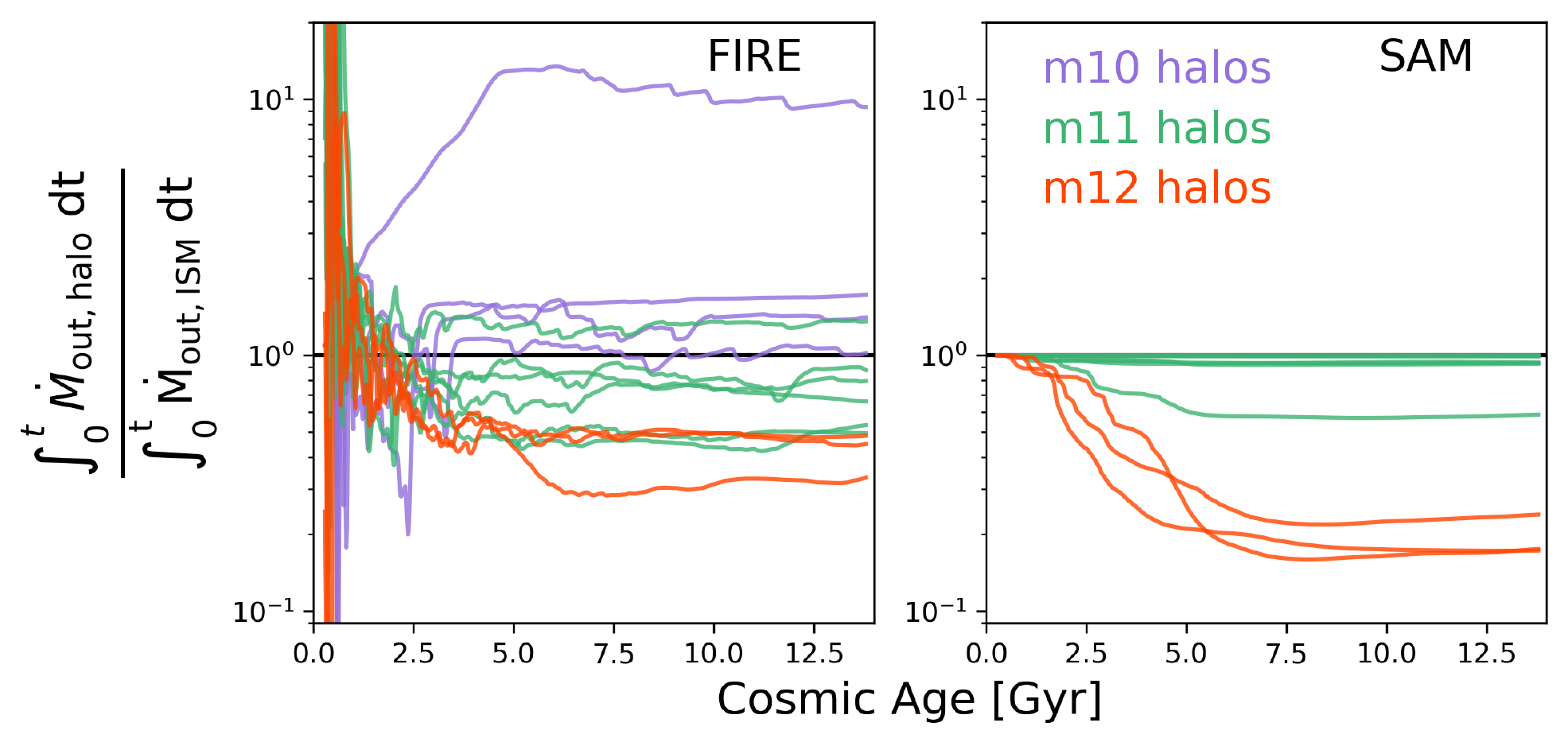}
\end{center}
\caption{The ratio of the cumulative mass ejected from the halo versus from the ISM as a function of cosmic time for FIRE-2 (left panel) and the SAM (right panel). The lines are color-coded according to $z=0$ halo mass. The SAM qualitatively follows the FIRE-2 trends for the m11 and m12 halos at $z\lesssim2$ (i.e., cosmic ages $\gtrsim$ 3 Gyr): an increasingly larger fraction of winds are able to leave the halo in progressively lower mass halos. But in the FIRE-2 simulations, the m10 halos strikingly tend to have expelled more material through $R_{\rm vir}$ than has ever left the ISM boundary ($0.1R_{\rm vir}$), implying significant entrainment of ambient CGM material by the outflows (this is also true for the progenitors of all halos at very early times $z\gtrsim6$). Since the ratio cannot exceed 1 in the SAM by construction, it asymptotes to 1 for the low-mass dwarfs (all of their winds leave the halo).}
\label{fig:sumout}
\end{figure*}

Lastly, we compare halo mass outflow rates in \autoref{fig:mdot_out_halo}. Again, the halo outflow rates are higher in the SAM than FIRE-2 for the m11 and m10 halos, and for the m12s at very early and at late times. This is somewhat expected given that the ISM outflow rates were higher as well, and the halo outflow rate is simply a halo circular velocity dependent re-scaling of the ISM outflow rate (\autoref{eqn:feject}). However, comparing the cumulative mass ejected from the halo versus from the ISM (obtained via integration of the respective mass outflow rate histories without boxcar smoothing) as a function of time between the two models reveals a striking phenomenon. In \autoref{fig:sumout}, we see that the ratio of halo outflow mass divided by ISM outflow mass is generally less than one in FIRE-2 for the m11 and m12 halos, except at very early times when the progenitor halos are in the dwarf phase. The SAM shows a qualitatively similar trend for these intermediate-mass dwarf and MW-mass halos at $z\lesssim2$: an increasingly greater fraction of wind mass is able to leave the halo in progressively lower mass halos. However, it is striking that in FIRE-2, this ratio can exceed 1 for the m10 dwarfs. The ratio reaches a factor of $\sim1.5$ for m10q and, incredibly, a factor of $\sim10$ for the late-forming m10v (and even higher ratios are reached for the progenitors of all halos at very early times $z\gtrsim6$). This implies that more mass has left the halo than has ever left the ISM (cumulatively), and is suggestive of entrainment of ambient CGM material by outflows \citep[see also][]{muratov15,hafen19,hafen20}. In contrast, the ratio can never exceed 1 by construction in the SC SAM. Hence, the SAM predicts that nearly all winds will leave the halo in low-mass dwarfs as specified by the function \autoref{eqn:feject}, but any potential effects resulting from entrainment are not captured by the SAM.

\section{Discussion}\label{sec:discussion}
Here we interpret the results from our comparison, discuss possible solutions to the model discrepancies with an emphasis on developing ways to modify SAMs to produce better agreement with FIRE, and present a new preventative stellar feedback model to help interpret the suppressed dwarf halo gas accretion efficiencies in FIRE-2.

\subsection{Interpreting the model discrepancies}\label{sec:diagnosis}
\begin{figure*}[!htbp] 
\begin{center}
\includegraphics[width=0.8\hsize]{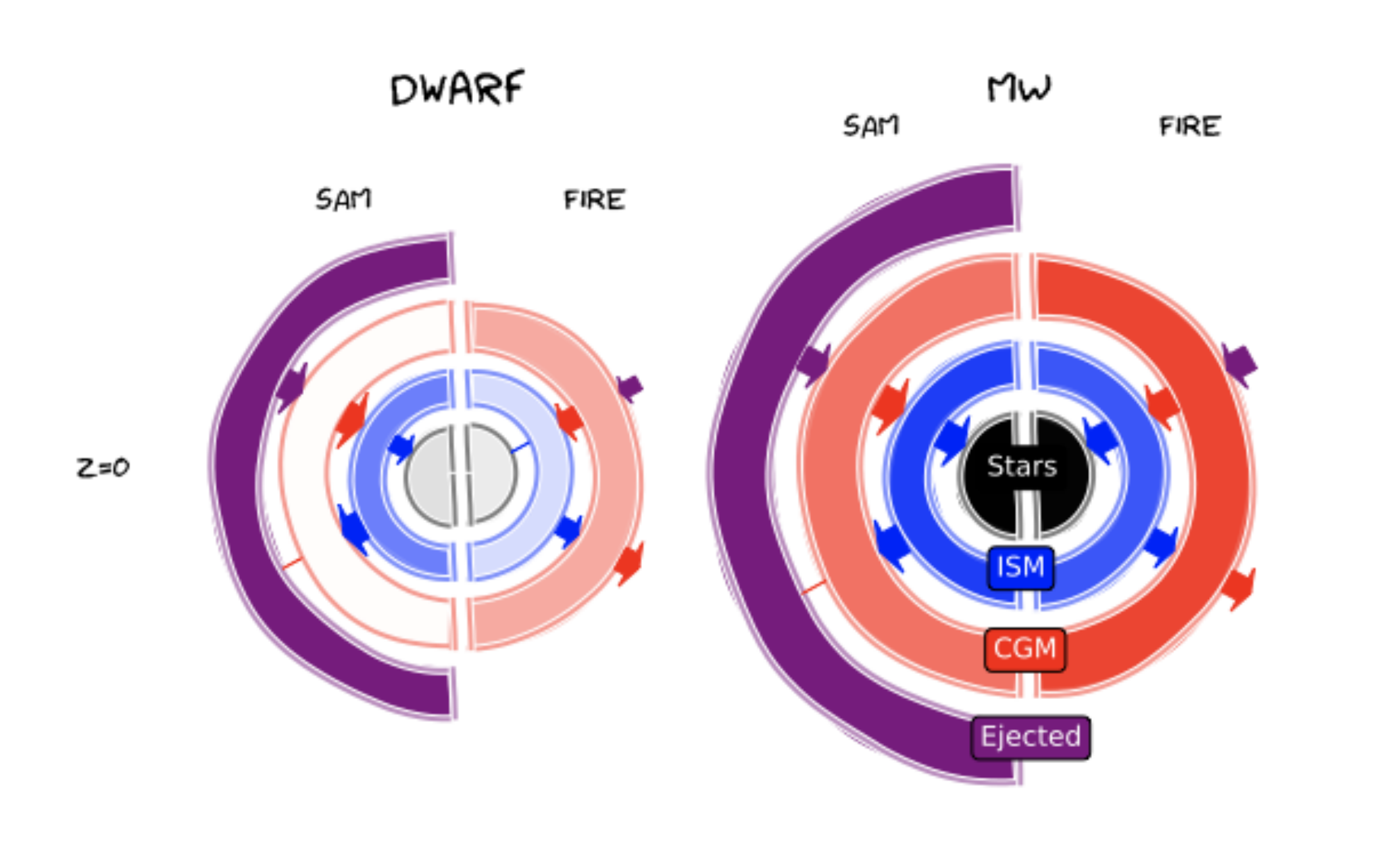}
\end{center}
 \caption{A cartoon that schematically illustrates the results of our comparison between the SAM and FIRE-2 (restricted to $z=0$ for simplicity). On the left we show a representative dwarf (m10q) and on the right a representative MW-mass halo (m12f). For each halo, the left half portrays the SAM and the right half depicts FIRE-2. From inside out, we show the bulk masses of stars (black), ISM (blue), CGM (red) and the ejected gas reservoir (magenta; restricted to the SAM since there is no clear definition of this component for FIRE-2). The opacity can be used to compare the mass of a single component between the two models or the mass of different components within a single model. The arrows illustrate inflows and outflows between the different bulk mass components (note that the purple arrows show the total halo gas accretion rate, not the recycling rate). Larger size arrows convey higher flow rates. Note how the stellar masses agree very well between the two models for both galaxies despite significant differences in the other bulk components.}
\label{fig:cartoon}
\end{figure*}

In this subsection, we will step through each of the four mass flow rates in turn and discuss the possible causes and solutions for the SAM versus FIRE-2 model discrepancies. But first we provide a high-level summary of the basic story, which is also summarized with a cartoon schematic in \autoref{fig:cartoon}. We showed that two very reasonable models of galaxy formation -- the Santa Cruz SAM and the FIRE-2 simulations -- agree relatively well with each other in terms of their stellar and ISM mass histories (\autoref{fig:obs}, \autoref{fig:mass_stars} and \autoref{fig:mass_ism}). However, the two models disagree dramatically in terms of their CGM mass histories, with the SAM remarkably predicting $\sim3$ orders of magnitude lower CGM mass than FIRE-2 for the dwarf halos (\autoref{fig:obs} and \autoref{fig:mass_cgm}). The SAM assumes that most of the ``missing" extragalactic gas resides outside of the halo in a so-called ``ejected" reservoir (owing partially to observational uncertainties about the total CGM masses of galaxies). To better understand the discrepancies, we turned to the actual mass flow rates for the ISM and CGM. The fundamental discrepancy between the SAM and FIRE-2 arises in the halo gas accretion rate (\autoref{fig:mdot_in_halo}). While there is reasonable agreement for the MW-mass halos, the SAM predicts much higher halo gas accretion rates for the dwarfs than FIRE-2 (exceeding a factor of ten for the low-mass dwarfs by $z=0$). The ISM gas accretion rates are also higher in the SAM than FIRE-2 by more than a factor of two for the m11 and m12 halos, and by more than a factor of ten for the low-mass dwarfs (\autoref{fig:mdot_in_ism}). These higher inflow rates in the SAM are compensated for by higher ISM and halo outflow rates in the SAM compared to FIRE-2 (\autoref{fig:mdot_out_ism} and \autoref{fig:mdot_out_halo}), making it possible to understand why the SAM and FIRE-2 predict similar stellar mass and ISM mass histories. In addition to these discrepancies, the SAM also does not capture star formation stochasticity (\autoref{fig:mdot_sfr}) and the entrainment of ambient CGM material by outflows from low-mass dwarfs (\autoref{fig:sumout}).

We begin by diagnosing the higher halo gas accretion rates of dwarf halos in the SAM compared to FIRE-2. We showed in \autoref{fig:mdot_in_halo} that re-accretion of previously ejected gas dominates over pristine accretion for the SAM dwarf halos. Hence, to first order the halo gas ejection and recycling model must be updated, but this is an area of uncertainty that has long plagued SAMs. Previous works have shown that the way in which halo gas ejection and re-accretion is implemented in SAMs can significantly affect results. Early models were split between allowing no re-accretion at all versus assuming a single re-accretion timescale \citep[e.g., see section 2.6 of][and references therein]{somerville99}. \citet{somerville08} claim that some re-accretion is necessary to match the observed baryon fractions of galaxy clusters (which would otherwise be predicted to be too low), but simultaneously reproducing the late formation times and mass functions of dwarfs has presented challenges. \citet{henriques13} proposed that the re-accretion timescale should depend inversely on halo mass with no dependence on redshift because that allowed their SAM to better match the observed evolution of dwarfs. \citet{white15} re-visited this issue with the Santa Cruz SAM and tested three alternative solutions for de-coupling the star formation and halo gas accretion histories of dwarfs: adding a redshift dependence for the mass loading factor of stellar-driven winds, changing the gas depletion timescale for star formation, and changing the re-accretion timescale as in \citet{henriques13}. Their comprehensive observationally-driven study concluded, in qualitative agreement with \citet{henriques13}, that preferentially increasing the re-accretion timescale for dwarfs may be the most promising solution.

Another approach for guiding SAMs is to explicitly track halo gas recycling in high-resolution simulations. \citet{anglesalcazar17} showed that recycling is ubiquitous and occurs over a broad range of timescales in the FIRE-1 simulations, although the recycling events generally happen in the inner halo (``fountain flows") leading to median recycling timescales of only $\sim100-350$ Myr \citep[see also][for the FIRE-2 suite]{hafen19,hafen20}. Interestingly, \citet{christensen16} and \citet{tollet19} both find longer median recycling timescales of $\sim1$ Gyr in their respective zoom-in simulation suites, with little or no dependence on halo mass, supporting the use of a single recycling time as adopted in some SAMs. Indeed, our SAM assumes a single recycling time (roughly on the order of a Hubble time), with the caveat that our recycling refers to gas already ejected from the halo whereas many of the previous simulation analyses define recycling within the halo. \citet{tollet19} and \citet{hafen20} also emphasize the inherently multi-phase nature of outflows in their simulations, with the hot component more easily able to leave the halo and the cooler component likely to be recycled at the inner halo via fountain flows. This general multi-phase picture for outflows is in agreement with even higher resolution but smaller scale simulations \citep[e.g.,][]{kim18,fielding18,li19a,li19b,kim20}. This suggests that decreasing the high halo gas accretion rates of dwarfs involves more than just preferentially increasing their recycling timescales, namely improving how the ejected SAM component is modeled in the first place. Some fraction of gas in the ejected reservoir should be allowed to become unbound permanently (especially for the hottest, fastest winds in dwarfs), and a distinct bound wind component should be modeled which does not have enough energy to escape beyond $R_{\rm vir}$ but instead may recycle rapidly in the inner halo. In addition, it may not be necessary to appeal solely to moving gas outside of the halo to reduce CGM cooling rates and hence SFRs; instead, some fraction of the ``ejected" reservoir could actually be placed within the CGM but in a thermal state that simply does not cool efficiently.

\autoref{fig:mdot_in_halo} also implies that updating halo gas ejection and recycling may not be enough: the pristine gas accretion rates alone can still be significantly higher in the SAM than FIRE-2 for the low-mass dwarfs (m10q is representative). It is tempting to attribute this to the different UV background model assumed in the SAM \citep[taken from the simulations of][who themselves adopted the UV background model of \citealt{haardt01}]{okamoto08} versus in FIRE-2 \citep{fauchergiguere09}. According to the \citet{okamoto08} prescription, the characteristic mass at which the bulk halo baryon fraction drops to half of the universal value at $z=0$ is $\approx10^{10}M_{\odot}$. It is a factor of a few higher, $\approx5\times10^{10}M_{\odot}$ according to the simulations of \citet[][their Figure 7]{fauchergiguere11}, who implemented the \citet{fauchergiguere09} UV background model.\footnote{But note that the \citet{fauchergiguere11} simulations pre-date the FIRE-2 subgrid models, hydrodynamic solver, etc. To properly assess the redshift evolution of the characteristic mass in FIRE-2, we would need FIRE-2 simulations with all stellar feedback turned off, such that only the UV background and gravitational shock heating can systematically suppress gas inflows at $R_{\rm vir}$.} Using the SAM, we have experimented with increasing the characteristic mass and/or changing the redshift of reionization (within a reasonable range of values) but find that this cannot satisfactorily explain the suppressed halo accretion, especially for the intermediate-mass dwarfs.\footnote{Note also that significantly changing the redshift of reionization or characteristic mass normalization would cause other predictions of the SAM to disagree with FIRE-2 and observations, and possibly make the SAM assumptions inconsistent with cosmology constraints from \citet{planck16}.} The main exception is m10v, the late-forming low-mass dwarf for which our SAM predictions disagree dramatically with FIRE-2: its virial mass does not exceed the SAM fiducial characteristic mass until $z\sim2$, compared to $z\sim10$ for the other m10 halos. Interestingly, previous authors have argued that the low halo baryon fractions and accretion rates of the FIRE-2 dwarfs can at least partially be attributed to the UV background \citep{fitts17,elbadry18,hafen19}, in contrast to the weaker effects predicted by the SAM.

What else could possibly suppress the halo gas accretion rates of the FIRE-2 dwarfs?\footnote{The large-scale cosmic web environment of a halo can be very relevant, but we think this is unimportant for our sample of FIRE-2 halos which are relatively isolated and have ``typical" accretion histories for their mass \citep{hopkins18}.} There is an emerging consensus that some form of preventative feedback is needed in SAMs beyond UV background heating alone. \citet{hirschmann12} already showed that the Santa Cruz SAM predicts much higher halo gas accretion rates compared to their reference suite of cosmological zoom-in simulations (see their Figure 11). Interestingly, \citet{lu11} found the opposite when comparing their SAM to the cosmological simulations of \citet{keres09}. Nevertheless, both of these authors later assumed general ``pre-heating" to suppress halo gas accretion rates for their SAMs \citep{lu15,hirschmann16,lu17}. More recently, \citet{tollet19} characterized the baryon cycle in the NIHAO simulations \citep{wang15} and also argued that SAMs would need a new ``maintenance feedback" channel to achieve lower cooling rates. They showed that in the NIHAO simulations, stellar-driven outflows from dwarf halos divert otherwise inflowing gas supplied by cosmic web filaments on scales as large as $6R_{\rm vir}$, resulting in suppressed accretion. Furthermore, the entrainment of outflows implied by our \autoref{fig:sumout} may have additional preventative feedback effects that need to be better understood \citep[this phenomenon is also seen in the FIRE-1, NIHAO and EAGLE simulations, respectively, by][]{muratov15,tollet19,mitchell19}. In the next section, we will present a simple but physically-motivated model for preventative stellar feedback that agrees remarkably well with the reduced halo gas accretion efficiencies in FIRE-2. 

Now we turn to the ISM inflow rate: the ISM accretion rates may be higher in the SAM than in FIRE-2 in part because the same is already true for the halo gas accretion rates. However, subtle details of the SAM halo gas cooling model \citep[based on][]{whitefrenk91} may also cause the ISM inflow rates to disagree. It is notable that although the CGM mass in the SAM is much lower than it is in FIRE-2 dwarfs, the ISM accretion rates are \emph{higher}. In the regime where $R_{\rm cool}<R_{\rm vir}$ (``hot/slow'' mode accretion), a higher overall CGM mass would likely correspond to higher cooling rates in the SAM (\autoref{sec:coolingmodel}). This implies that if the SAM CGM masses were somehow made to agree better with FIRE-2, the ISM inflow discrepancy would presumably become worse with the existing SAM cooling model. However, the simple SAM assumption that gas accretes into the ISM on a dynamical time when $R_{\rm cool}>R_{\rm vir}$ (``cold/fast" mode accretion) could also be a factor. If gas accretes into the ISM too quickly, without spending enough time in the CGM, this would be consistent with both the lower CGM masses and higher ISM inflow rates of SAM dwarfs compared to FIRE-2. Since it is likely that most of the dwarf halos spend most of their lifetime experiencing this so-called ``cold/fast" mode accretion in the SAM, this is an important regime to study in the future.

A critical point is that the SAM does not include a heating term due to stellar-driven winds that can offset the predicted halo gas cooling rate. In the FIRE-2 dwarfs, it is almost certainly the case that the energy and momentum of stellar-driven winds are suppressing accretion on the scale of the ISM (this may even have an effect in the MW-mass halos at late times, where the SAM ISM accretion rates are higher than FIRE-2 by more than a factor of two; \autoref{fig:mdot_in_ism}). In addition, the calculation of a ``cooling radius" and ad hoc treatment of the case when it is greater than the virial radius can lead to unphysical looking behavior for the dwarfs (e.g., the ``boxy" trajectories for CGM mass in \autoref{fig:mass_cgm}). Even for the radiative cooling timescale calculation itself, the assumed singular isothermal CGM mass density profile is likely an oversimplification for the simulated halos since bursty inflows and outflows may cause the CGM to have a more dynamic structure. SAMs do not generally explicitly model the structure and dynamics of the CGM, but this is slowly changing with work on new cooling flow solutions \citep[e.g.,][]{lu11,stern19} and explicit CGM substructure models \citep[e.g.,][]{maller04,voit15,faerman19,lan19}. Explicitly modeling the CGM with SAMs is important given that modern cosmological hydrodynamical zoom-in simulations, including the FIRE suite, might lack the resolution requirements to capture some of the relevant cooling and shock heating microphysics \citep[e.g., see the recent enhanced halo resolution studies by][]{hummels19,vandevoort19,peeples19}. 

Finally, switching to the outflow side: it is again not surprising that the ISM outflow rates are much higher in the SAM than in FIRE-2 given the agreement between their stellar mass histories. The only plausible way to decrease the SAM ISM outflow rates is to implement preventative feedback that suppresses the high gas accretion rates in the first place. Improvements in this area may fundamentally require changing how we model ``disk mode" star formation and what we assume about variations in the local star formation efficiency \citep[e.g.,][and references therein]{khullar19}. Indeed, the order of magnitude ISM mass excess but factor of two agreement on stellar mass for low-mass dwarfs predicted by the SAM compared to FIRE-2 suggests that the
assumptions for how gas forms stars are different in the two models. In addition, small-scale simulations suggest that preventative feedback effects may be stronger during bursty star formation episodes since those result in clustered supernovae that drive faster, more energetic winds \citep[e.g.,][]{gentry17,fielding17}. To achieve local star formation efficiency variations and stochasticity in a physically self-consistent way, the SAM outflow model itself may need to be replaced with one that depends exclusively on local properties. Ideally, on the ISM scale, the mass, energy, momentum and metal mass from stellar feedback should be deposited locally, e.g., within annuli of a radially-resolved disk \citep[e.g.][]{forbes19}. For halo outflows, while the traditional approach of setting a wind escape fraction that depends on the global halo circular velocity may still be viable, \autoref{fig:sumout} suggests a need for additional variables that account for entrainment and ejection of ambient CGM material by multi-phase outflows \citep[see also][]{guo11,muratov15,hu19,li19a,tollet19,mitchell19,hafen19,hafen20}.

The discrepancies between the Santa Cruz SAM and FIRE-2 have implications for other models of galaxy formation. That two models with very different underlying baryon cycles can still match the observed evolution of the stellar mass function, and by extension the low-mass end of the stellar-to-halo mass relation, emphasizes ambiguities for interpreting observations with phenomenological models. These ambiguities are amplified even more with subhalo abundance matching and ``semi-empirical models" that make even simpler assumptions for how halo mass accretion rates relate to galaxy star formation rates \citep[e.g.,][]{behroozi13c,moster13,rodriguezpuebla16,rodriguezpuebla17,moster18,tacchella18,behroozi19}. For example, it is common practice in these models to define the star formation efficiency as SFE=$\frac{\rm SFR}{f_{\rm b}\dot{M}_{\rm vir}}$. If indeed the halo gas accretion rates of dwarfs follow the universal baryon fraction, then this would imply low SFEs in dwarfs. But our study suggests that it is also possible for the reverse interpretation to be true: for a given SFE, if less gas is flowing into the halo in the first place, then this can also explain the lower SFRs of dwarfs. With a SAM coupled to high resolution simulations, we can explicitly isolate and model these preventative physical processes (as in the next section) and ultimately study the implications for the evolution of the galaxy--halo connection.

\subsection{A simple preventative stellar feedback model}\label{sec:preventative}
Here we present a simple physical model for how preventative stellar feedback can suppress gas accretion rates preferentially for dwarf halos on the scale of $R_{\rm vir}$.  We deliberately keep the model simple as the goal here is to demonstrate that a reasonable model can approximately match the FIRE-2 inflow results, rather than trying to develop a detailed prescription for inclusion in semi-analytic models, a task we defer to future work. The essence of our model is that the energy from SN-driven winds can heat some fraction of the gas beyond $R_{\rm vir}$ to the virial temperature (or higher). Since the virial temperature is a measure of the gravitational potential depth, this would then imply that the heated gas becomes unbound from the halo and hence is unable to accrete.\footnote{The unbound, low-density hot gas may then travel outwards before eventually turning around and recooling on to the halo \citep[e.g., as illustrated in Figure 1 of][]{noh14}. More complicated models may predict the detailed evolution of this gas, but here we restrict ourselves to simply deriving an effective suppression fraction for the initially accreting gas.} Note that preventative feedback in this context refers to preventing gas from accreting into the halo in the first place \citep[as in][]{lu17}, rather than preventing halo gas from accreting into the ISM \citep[e.g.,][]{mitra15}.

First, we define 
\begin{equation}
f_{\rm in} = \frac{\dot{M}_{\rm in,baryons}}{f_{\rm b}\dot{M}_{\rm in,DM}}
\end{equation}
as the ratio of the actual baryonic mass inflow rate ($\dot{M}_{\rm in,baryons}$) to the baryon fraction-adjusted DM mass inflow rate ($f_{\rm b}\dot{M}_{\rm in,DM}$) at the virial radius.\footnote{Note that we should also multiply by an additional factor $f_{\rm coll}$ to account for UV background heating (\autoref{eqn:pristine}), but this is likely negligible for most of our halo masses, as we will show later.} 

We can obtain an expression for the amount of gas mass that must be heated to suppress the accretion rate by first writing down an expression for halo gas binding energy:
\begin{equation}
E_{\rm b} = \frac{1}{2} f_{\rm b} M_{\rm vir} V_{\rm vir}^2\;.
\end{equation}
Next, we take the time derivative of this expression and equate it to the heating rate ($\dot{E}_{\rm heat}$); if we assume $V_{\rm vir}$ is constant and we isolate the gas mass term ($f_{\rm b}M_{\rm vir}$), we get the needed mass heating/unbinding rate: 
\begin{equation}
f_{\rm b} \dot{M}_{\rm in,DM} (1-f_{\rm in}) = \frac{2\dot{E}_{\rm heat}}{V_{\rm vir}^2}\;,
\end{equation}
where we have used our definition of $f_{\rm in}$ to replace $\dot{M}_{\rm in,baryons}$. The factor $(1-f_{\rm in})$ comes in because we want to equate the heating rate with the fraction of gas that does not accrete.\footnote{An alternative derivation is to directly balance the heating rate with the specific gravitational potential energy of the fraction of gas beyond $R_{\rm vir}$ that was heated to at least $T_{\rm vir}$ and hence unable to accrete: $\dot{E}_{\rm heat} = f_{\rm b} \dot{M}_{\rm in,DM} (1-f_{\rm in}) \frac{k_B T_{\rm vir}}{\mu m_H}$.}

Assuming the heating is provided by energy from star formation, we can write $\dot{E}_{\rm heat} = \eta_{\rm E} e_{\rm SN}\rm SFR$, where SFR is the star formation rate, $e_{\rm SN}=10^{51}\textrm{erg}/(100M_{\odot})$ is the specific energy produced by SNe per $100 M_{\odot}$ of stars formed (this is approximate at the order of magnitude level given a reasonable assumption for the IMF), and $\eta_{\rm E}$ is the efficiency in transporting that energy from the SN site to the virial radius. Doing this allows us to solve for $f_{\rm in}$:
\begin{equation}
f_{\rm in} = 1 - 2\eta_{\rm E} \frac{e_{\rm SN}}{V_{\rm vir}^2}   \frac{\rm SFR}{f_{\rm b} \dot{M}_{\rm in,DM}}\;.
\label{eq:model-fin}
\end{equation}

Note that the ratio $e_{\rm SN}/V_{\rm vir}^2$ will be higher for dwarfs owing to their lower $V_{\rm vir}$. The other important term is the star formation efficiency ratio SFR/$(f_{\rm b} \dot{M}_{\rm in,DM})$. Since the SFE defined in this way is generally lower for dwarfs, this new term acts in the opposite direction of the $e_{\rm SN}/V_{\rm vir}^2$ trend. To make further progress, we therefore need a prediction for SFR (or equivalently SFE). One option is to take this from the SAM or simulation itself (perhaps suitably time-shifted to allow for a delay as the energy flows from the ISM to the virial radius). However, here we assume a simple equilibrium ``bathtub" model \citep[e.g.,][]{dave12,mitra15} in which the amount of gas in the ISM is fixed (at least over short periods of time) such that the amount of inflowing gas is balanced by the outflowing gas. In this case, we can write
\begin{equation}
    \textrm{SFR} = \frac{f_{\rm in} f_{\rm b} \dot{M}_{\rm in,DM}}{1 + \eta_{\rm M}},
    \label{eq:model-SFR}
\end{equation}
where $\eta_{\rm M}$ is the mass-loading factor, or ratio of the mass outflow rate (near the ISM) to the star formation rate. Using this relation in \autoref{eq:model-fin} allows us to solve for $f_{\rm in}$:
\begin{equation}\label{eqn:main}
f_{\rm in} = (1+\psi)^{-1}\,,
\end{equation}
where 
\begin{equation}
\psi \equiv \frac{2\eta_{\rm E} e_{\rm SN}}{(1 + \eta_{\rm M}) V_{\rm vir}^2} 
\end{equation}
is the ratio of specific SN energy to the specific halo gravitational potential, accounting for our mass and energy loading efficiencies. If the ratio $\eta_{\rm E}/(1+\eta_{\rm M})$ was a constant, then $\psi$ will be larger and hence $f_{\rm in}$ will be smaller for lower mass halos. This would give the expected qualitative behavior that a lower fraction of gas is able to accrete into dwarf halos. However, as a last step, we need to explicitly consider how $\eta_{\rm M}$ and $\eta_{\rm E}$ may evolve with halo mass and/or redshift. For $\eta_{\rm M}$, we directly take the broken power law relation for the FIRE-1 simulations from \citet{muratov15}. According to their equations 4 and 5, $\eta_{\rm M}$ follows a steeper power law for halos with $V_{\rm vir}<60$ km s$^{-1}$ and there is a slight redshift dependence. For the energy loading factor $\eta_{\rm E}$ on the scale of $R_{\rm vir}$, there is less precedent. We therefore consider two simple possibilities. First, we assume a constant $\eta_{\rm E}=0.1$ motivated by the ISM wind breakout condition study of \citet{li19a}. Alternatively, we hypothesize that lower mass halos have preferentially higher energy loading factors \citep[which is plausible given their preferentially higher mass loading factors and the apparently energy-conserving nature of their winds;][]{muratov15}. Specifically, we assume $\eta_{\rm E}=\varepsilon_{\rm heat}(1+\eta_{\rm M})$ where $\varepsilon_{\rm heat}$ is a constant that parameterizes our ignorance about the conversion from ISM mass loading to ISM energy loading and then to halo energy loading. With this simple parameterization, the strong halo mass (and slight redshift) dependence of $\eta_{\rm M}$ from \citet{muratov15} is canceled out, allowing us to see how our model behaves if indeed the ratio $\eta_{\rm E}/(1+\eta_{\rm M})$, rather than $\eta_{\rm E}$ alone, is constant.

\begin{figure*}[!htbp] 
\includegraphics[width=0.8\hsize]{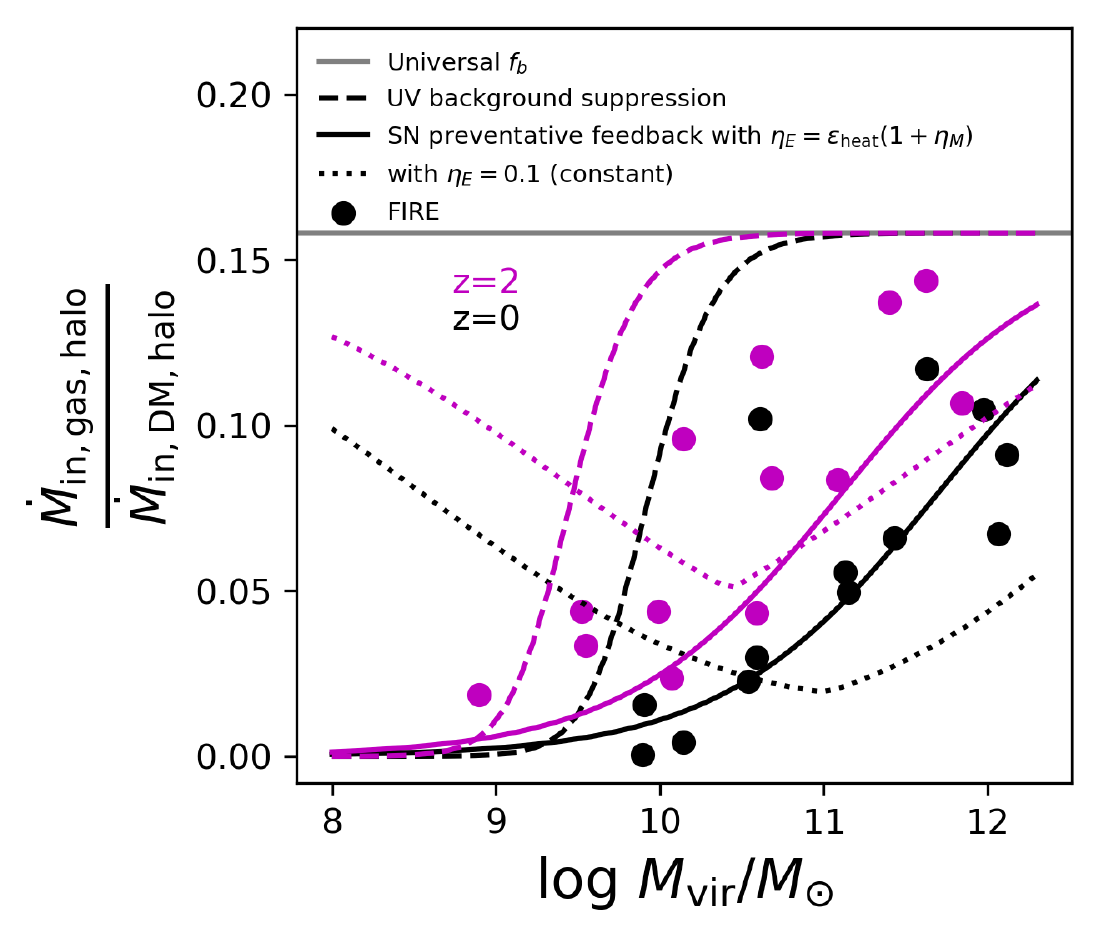}
 \caption{The halo gas accretion efficiency ($\dot{M}_{\rm in,gas}/\dot{M}_{\rm in,DM}$ in the virial shell) as a function of the halo virial mass for the FIRE-2 simulations at $z=0$ (black points) and $z=2$ (magenta points). If gas accretion perfectly tracked DM accretion at $R_{\rm vir}$ with the universal baryon fraction \citep[$f_{\rm b}=0.158$ from][]{planck16} as commonly assumed, then the halos would all lie along the horizontal solid gray line. The \citet{okamoto08} model describing the suppression of halo gas accretion due to the ionizing UV background is shown with the black dashed line for $z=0$ and the magenta dashed line for $z=2$. The FIRE-2 halo gas accretion efficiencies fall below the expectation from UV background heating alone. The dotted lines show the behavior of our simple preventative stellar feedback model if we assume the halo energy loading factor is a constant $0.1$; we see that it is incapable of explaining the data points, in fact showing the opposite trend at low masses. However, the solid lines show that our model can explain the data points remarkably well if we assume that the halo energy loading factor is preferentially higher for dwarfs ($\eta_{\rm E}=\varepsilon_{\rm heat}\eta_{\rm M}$ with $\varepsilon_{\rm heat}=0.01$, implying that $\eta_{\rm E}$ is of order unity for low-mass dwarfs). The agreement is better at $z=0$ than $z=2$, suggesting either a redshift dependence for $\varepsilon_{\rm heat}$ or that our simple model is breaking down.}
\label{fig:fb}
\end{figure*}

In \autoref{fig:fb}, we plot the halo gas accretion efficiency as a function of halo mass for the FIRE-2 halos at $z=0$ and $z=2$, where we define the accretion efficiency to be the ratio of the gas accretion rate to the DM accretion rate in the virial shell (without any boxcar time smoothing). If gas accretion perfectly tracked DM accretion at $R_{\rm vir}$ as commonly assumed in halo models, then the halos should all lie along the mass-independent universal baryon fraction line \citep[$f_{\rm b}=0.158$;][]{planck16}. We know that heating from the UV background can preferentially suppress gas accretion into low-mass halos, so this cannot be strictly true. However, as we have already discussed above, the accretion efficiencies of the FIRE-2 halos fall below the expected suppression due to the UV background alone \citep[comparing to][which is the relation assumed in the Santa Cruz SAM]{okamoto08}. This is perhaps not so surprising because UV background heating is thought to primarily affect halos with much lower masses than ours. Turning to the version of our model with a constant $\eta_{\rm E}=0.1$, we see that it is incapable of describing the data points; in fact, at low halo masses, this version of the model shows an upturn in $f_{\in}$. However, if we adopt the second version of the model with $\eta_{\rm E}=\varepsilon_{\rm heat}(1+\eta_{\rm M})$, and set $\varepsilon_{\rm heat}=0.01$ (implying that $\eta_{\rm E}$ is preferentially higher, i.e., of order unity, in the low-mass dwarfs), then the prediction from our simple model matches the data points remarkably well, especially at $z=0$. The predicted suppression at $z=2$ is somewhat stronger than the data points, which may suggest that $\varepsilon_{\rm heat}$ should have a redshift dependence and/or that our simple equilibrium bathtub model is breaking down.

We again stress that our preventative stellar feedback model is very simple and although it is promising, there are several unknowns that should be addressed in the future. First and foremost, we started by assuming that SN-driven winds can reach $R_{\rm vir}$ and heat a fraction of the surrounding gas to the virial temperature or higher. This is certainly a plausible assumption for the FIRE-2 dwarfs given their high halo-to-ISM cumulative outflow mass ratios (\autoref{fig:sumout}). It should be less the case in the MW-mass halos since winds would need a higher velocity to escape the potential well of these more massive halos; however, in detail this depends on the relative fraction of hot, fast-moving wind versus cooler, slower moving wind, and the rate at which the thermal energy of the wind is lost to the ambient CGM due to interactions/mixing (we have not distinguished between kinetic and thermal energy for the SN winds). Directly characterizing $\eta_{\rm E}$ and $\eta_{\rm M}$ for the FIRE-2 halos would be of great interest for testing and calibrating our model in the future. In addition, there will be degeneracies between preventative stellar feedback, ejective stellar feedback, and gravitational shock heating of gas accreting onto the more massive halos. The implications of these degeneracies for the galaxy--halo connection can be explored with a SAM in the future, provided that the physical processes have been modeled and calibrated to faithfully represent the hydrodynamical simulations.

Many previous works have already suggested that preventative stellar feedback is important in dwarfs. \citet{dekel86} derived the equations for SN-driven heating of halo gas and the implications for ejecting gas (based on comparing the specific SN energy with the halo virial temperature; see their section 4). Here we are explicitly considering suppression of gas accretion rather than gas ejection alone. \citet{oppenheimer09} and \citet{oppenheimer10} used hydrodynamical simulations to infer that SN-driven winds must have an additional heating/preventative effect to offset gas cooling, but their results were not parameterized and easily translatable to SAMs \citep[see also][]{pawlik09,vandevoort11,christensen16,elbadry18,wright20}. \citet{lu15} and \citet{lu17} explicitly implemented preventative feedback in their SAM and found that it is required (along with ejective feedback) to simultaneously explain the observed stellar mass--metallicity relation and the stellar mass function. However, their preventative feedback equation is more schematic in nature, and can be ascribed to ``pre-heating" by a multitude of processes in a more general sense \citep[see also][]{hirschmann16}. In contrast, we have explicitly constructed a model that isolates one potentially important preventative effect of SN-driven winds alone. There are likely additional preventative stellar feedback effects such as an energy input rate into the ambient CGM that can offset the predicted radiative cooling rate and possibly even eject ambient CGM material \citep[e.g.,][]{guo11}.

\section{Summary}\label{sec:summary}
We have used the FIRE-2 cosmological hydrodynamical ``zoom-in" simulations \citep{hopkins18} to test some of the fundamental assumptions in the Santa Cruz SAM \citep{somerville15} related to the global baryon cycle. We ran the Santa Cruz SAM on the FIRE-2 merger trees and compared, on an individual halo-by-halo basis, the time evolution of the masses of various components (stars, ISM, CGM) and the corresponding mass flow rates into and out of the ISM and CGM. We did not change anything in the SAM (which has been shown capable of matching many observations at $z=0$ and higher redshift) except to turn off AGN feedback since that is not included in the FIRE-2 simulations we use. Our sample spans 13 halos grouped into three mass bins, with at least 3 halos per bin: low-mass dwarfs ($M_{\rm vir}\sim10^{10}M_{\odot}$ at $z=0$), intermediate-mass dwarfs ($M_{\rm vir}\sim10^{11}M_{\odot}$), and MW-mass galaxies ($M_{\rm vir}\sim10^{12}M_{\odot}$). We also presented a simple physical model for how preventative stellar feedback can suppress halo gas accretion on the scale of $R_{\rm vir}$ preferentially for dwarfs.

Our main takeaways are as follows: 

\begin{enumerate}
\item At $z=0$, the SAM agrees relatively well with FIRE-2 and empirical constraints on the stellar-to-halo mass relation. The SAM and FIRE-2 also agree relatively well with each other and with observations for the ISM-to-stellar mass ratio at $z=0$ (as a function of stellar mass). However, they disagree dramatically with each other in terms of CGM mass: the CGM mass of dwarfs is $\sim3-4$ orders of magnitude lower in the SAM than in FIRE-2. This reflects the flexibility allowed in galaxy formation models to match observations of stars and the ISM while at the same time disagreeing greatly on the total CGM mass (owing partially to the observational uncertainty about whether most extragalactic gas resides within or outside of halos). 

\item As a function of time, the SAM reproduces the stellar mass assembly histories of the FIRE-2 galaxies generally within a factor of two (with the exception of one late-forming dwarf m10v). However, despite the overall agreement on the stellar mass assembly history, the two models disagree on the star formation history on shorter timescales of $\sim100$ Myr. The SAM does not demonstrate stochasticity in its SFHs whereas it is an ubiquitous phenomenon in the FIRE-2 dwarfs at all times and in the FIRE-2 MW-mass halo progenitors at early times \citep[as also shown by][]{muratov15,anglesalcazar17,ma17,sparre17,fauchergiguere18}. 

\item The time series of ISM mass agrees relatively well between the SAM and FIRE-2, although the SAM tends to be higher than FIRE-2 for the low-mass dwarfs. The CGM mass discrepancy between the SAM and FIRE-2 at $z=0$ (at fixed mass) also extends over all time. The mass of the ``ejected" gas reservoir in the SAM dominates over the CGM mass at all times, even in the MW-mass halos; this previously ejected gas is assumed to re-accrete back into the CGM on roughly a Hubble time in the SAM. Despite these dramatic differences in the individual bulk components, the halo baryon fractions tend to agree within a factor of $\sim2-3$ at all times, and both the SAM and FIRE-2 show the same qualitative trend: lower mass halos are more depleted of baryons than higher mass halos, relative to the universal baryon fraction. 

\item Comparing the mass flow rates as a function of time gives clues to the discrepancies in the integrated masses. The fundamental mismatch is that the SAM has significantly higher halo gas accretion rates compared to FIRE-2, with the discrepancy being systematically worse for dwarfs by a factor of $\sim10-100$. We argue that this is due to a combination of high re-accretion rates of gas previously ejected from the halo and the lack of preventative stellar feedback to suppress pristine halo gas accretion. The ISM accretion rates are also higher in the SAM than in FIRE-2, owing primarily to the halo gas cooling model and lack of preventative stellar feedback in the SAM. Correspondingly, to match the stellar assembly histories and the observed $z=0$ stellar mass function, the SAM has higher mass outflow rates than FIRE-2 from both the ISM and the halo. But, the low-mass dwarfs in FIRE-2 have cumulatively ejected more mass from their halo than has ever left their ISM (between a factor of $\sim1.5$ up to $\sim10$ by $z=0$; even larger ratios are measured for the progenitors of all halos at very early times $z\gtrsim6$). This implies significant entrainment of ambient CGM material and may have important preventative feedback effects that are not currently allowed for in the SAM by construction.

\item We propose a simple physical model for how stellar-driven winds can suppress halo gas accretion on the scale of $R_{\rm vir}$ for dwarfs. The essence of the model is that SN-driven winds can shock heat some fraction of gas beyond $R_{\rm vir}$ to the virial temperature or higher such that it can no longer accrete into the halo. We show that this simple model is capable of reproducing the reduced halo gas accretion efficiencies of the FIRE-2 dwarfs remarkably well, provided that the energy loading factor at $R_{\rm vir}$ is preferentially higher for dwarfs. Characterizing the mass and energy loading factors from the simulations in the future will help test and calibrate our preventative stellar feedback model.

\end{enumerate}

Given all of the model discrepancies and potential improvements discussed herein, it is natural to ask whether a SAM that is calibrated to match FIRE-2 (or any zoom-in simulation suite) can also still match observations. This is one of our ultimate driving questions, but our work demonstrates that the overall foundational structure and perhaps philosophy underlying SAMs may first need to be updated. Explicitly adding preventative feedback is arguably the most crucial step because the current Santa Cruz SAM does not contain the relevant physics to capture the low halo gas accretion rates of FIRE-2 dwarfs. The apparent success of our new preventative stellar feedback model suggests a path forward, but automated parameter space exploration techniques will be needed to map out degeneracies with existing SAM assumptions. Beyond that, we will need to improve (among other things) how we model halo gas cooling, the multi-phase structure of the CGM, the stochastic nature of star formation and associated outflows, and implement new channels for halo gas ejection and recycling. In parallel, it will be important to consider the statistical challenges associated with calibrating a SAM using the relatively small sample size of halos that can be provided by modern zoom-in simulation suites, and then scaling up predictions to the level of galaxy populations. In particular, it is presently unclear if the diversity in halo growth histories at a fixed halo mass is enough on its own to reproduce the scatter in galaxy properties at a fixed stellar mass, or if there is additional scatter on the baryonic physics side (e.g., from smaller sub-galactic scales) that needs to be modeled. These and related issues will be the subject of our future work.

\acknowledgements
This work was carried out as part of the SMAUG Project. SMAUG gratefully acknowledges support from the Center for Computational Astrophysics at the Flatiron Institute, which is supported by the Simons Foundation. We thank Andrew Wetzel and Peter Behroozi for help with running the Rockstar halo finder and consistent-trees merger tree code on the FIRE-2 halos. We also thank Joel Primack, Sandy Faber, Kevin Bundy, Piero Madau, Eve Ostriker, Volker Springel, Yakov Faerman and Nir Mandelker for helpful conversations and comments. We thank the anonymous referee for helpful suggestions that improved the clarity of this paper. VP is supported by the National Science Foundation Graduate Research Fellowship Program under Grant No. 1339067. GLB acknowledges financial support from the NSF (grant AST-1615955, OAC-1835509), and NASA (grant NNX15AB20G), and computing support from NSF XSEDE. rss gratefully acknowledges support from the Simons Foundation. The work of C.-G.K. was in part supported by a grant from the Simons Foundation (CCA 528307, E.C.O.) and in part by NASA ATP grant No. NNX17AG26G. The data used in this work were, in part, hosted on facilities supported by the Scientific Computing Core at the Flatiron Institute, a division of the Simons Foundation.

\vspace*{5mm}
\software{yt \citep{yt}, Python \citep{python}, IPython \citep{ipython}, NumPy \citep{numpy}, SciPy \citep{scipy}, Matplotlib \citep{matplotlib}, H5py \citep{h5py}, AstroPy \citep{astropy1,astropy2}}

\bibliographystyle{aasjournal}
\bibliography{references}

 \appendix
 
\section{Results using dark matter only simulations}\label{sec:dmonly} 
Throughout this paper, we have run the SAM on merger trees extracted for dark matter halos from the main hydrodynamical FIRE-2 simulations, i.e., baryons have affected the properties of DM halos in the merger trees. This was done to increase our sample size of halos (13) because DM-only $N$-body simulations do not exist for all of the FIRE-2 halos. Here, we re-run the SAM on a subset of the FIRE-2 halos that have corresponding DM-only simulations available (with the same initial conditions, resolution, snapshot output times, etc.). We run Rockstar and consistent-trees on these DM-only simulations in the same way as described in \autoref{sec:analysis} except we no longer force Rockstar to up-weight the DM particle mass since there are no baryonic particles to account for.

\begin{figure*}[!htbp] 
 \begin{center}
 \includegraphics[width=0.9\hsize]{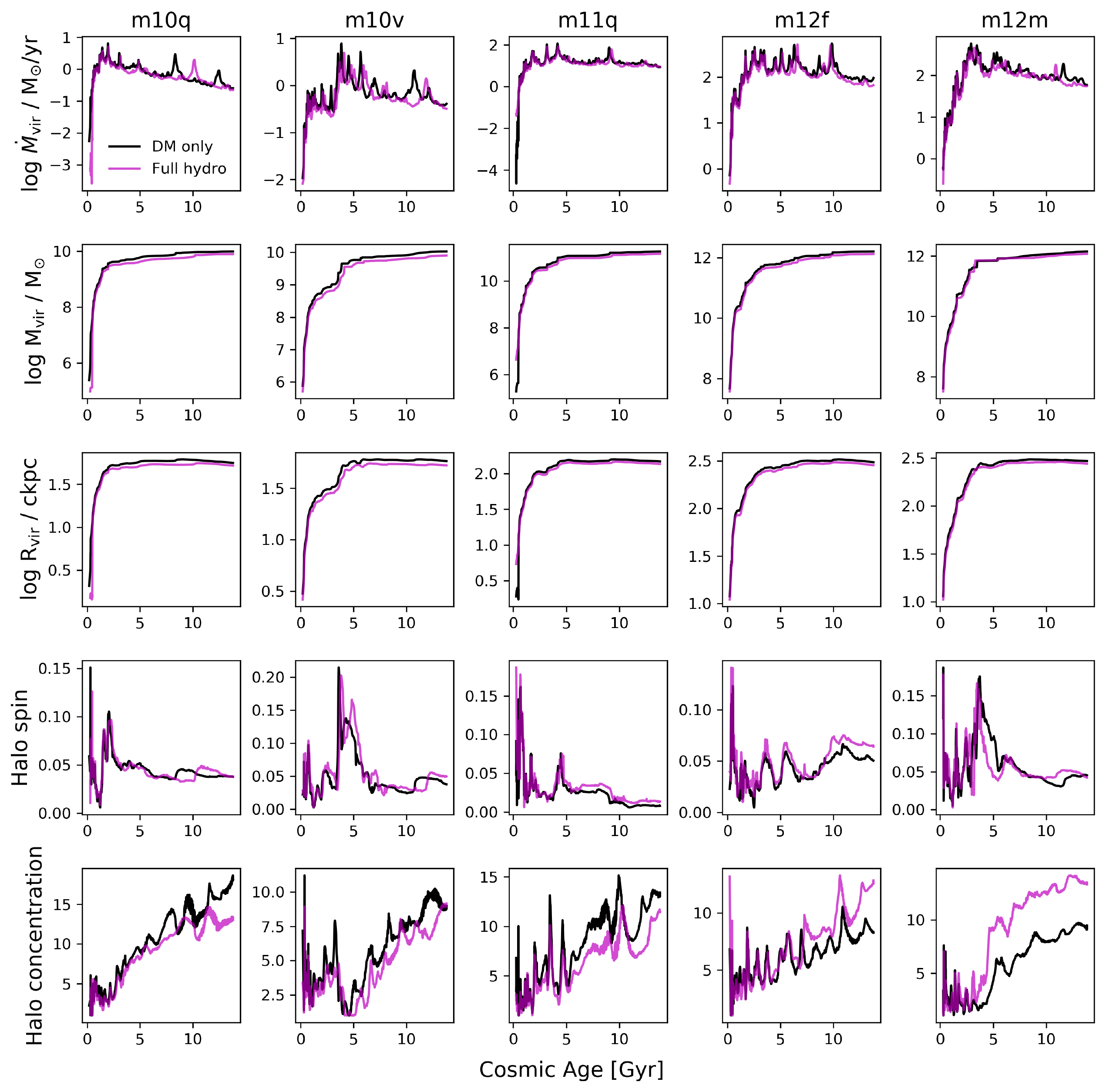}
 \end{center}\caption{A few relevant halo properties measured in the full hydrodynamical simulations (magenta lines) and the corresponding DM-only simulations (black lines). These are the 5 halos for which corresponding DM-only FIRE-2 runs exist. The DM-only halo properties are very similar to the hydro-based halo properties, with the Mvir and Rvir being lower in the hydro version by only 10-20\% on average. The major systematic difference is in the halo concentration which tends to be lower in the dwarfs in the hydro version (presumably due to stellar feedback adiabatically expanding the halo center) whereas it is higher for the MW-mass halos in the hydro run (presumably due to the significant stellar mass adiabatically contracting the halo center).}
 \label{fig:dmonly_fire}
 \end{figure*}

Before comparing the SAM results, it is useful to compare a few relevant halo properties measured with Rockstar in the hydrodynamical versus corresponding DM-only $N$-body simulations. \autoref{fig:dmonly_fire} overplots the time series of the halo DM mass accretion rate, $M_{\rm vir}$, $R_{\rm vir}$, halo spin parameter and halo concentration from the two matching runs for each of the five halos. The baryonic effects on these DM halo properties are generally not significant. The DM mass accretion histories have the same normalization on average, except that some spikes in the halo mass accretion rate are suppressed in the hydrodynamical version of the merger trees. This suppression of spikes in the mass accretion history is likely related to the virial mass and virial radius being smaller in the hydrodynamical run compared to the pure DM-only run, although the difference is only at the $10-20\%$ level. The halo spin parameters are nearly identical as a function of time. The main systematic trend is in the halo concentration parameter: lower mass halos have lower concentrations in the hydrodynamical run, presumably due to adiabatic expansion of the halo due to strong baryonic feedback \citep[][]{fitts17}. In contrast, the halo concentration parameter is larger in the hydrodynamical simulation for the MW-mass halos, presumably due to the greater central mass of baryons leading to adiabatic contraction of the halos \citep[as is analytically expected, e.g.,][]{mo98,dutton16}.

\begin{figure*}[!htbp] 
 \begin{center}
 \includegraphics[width=0.9\hsize]{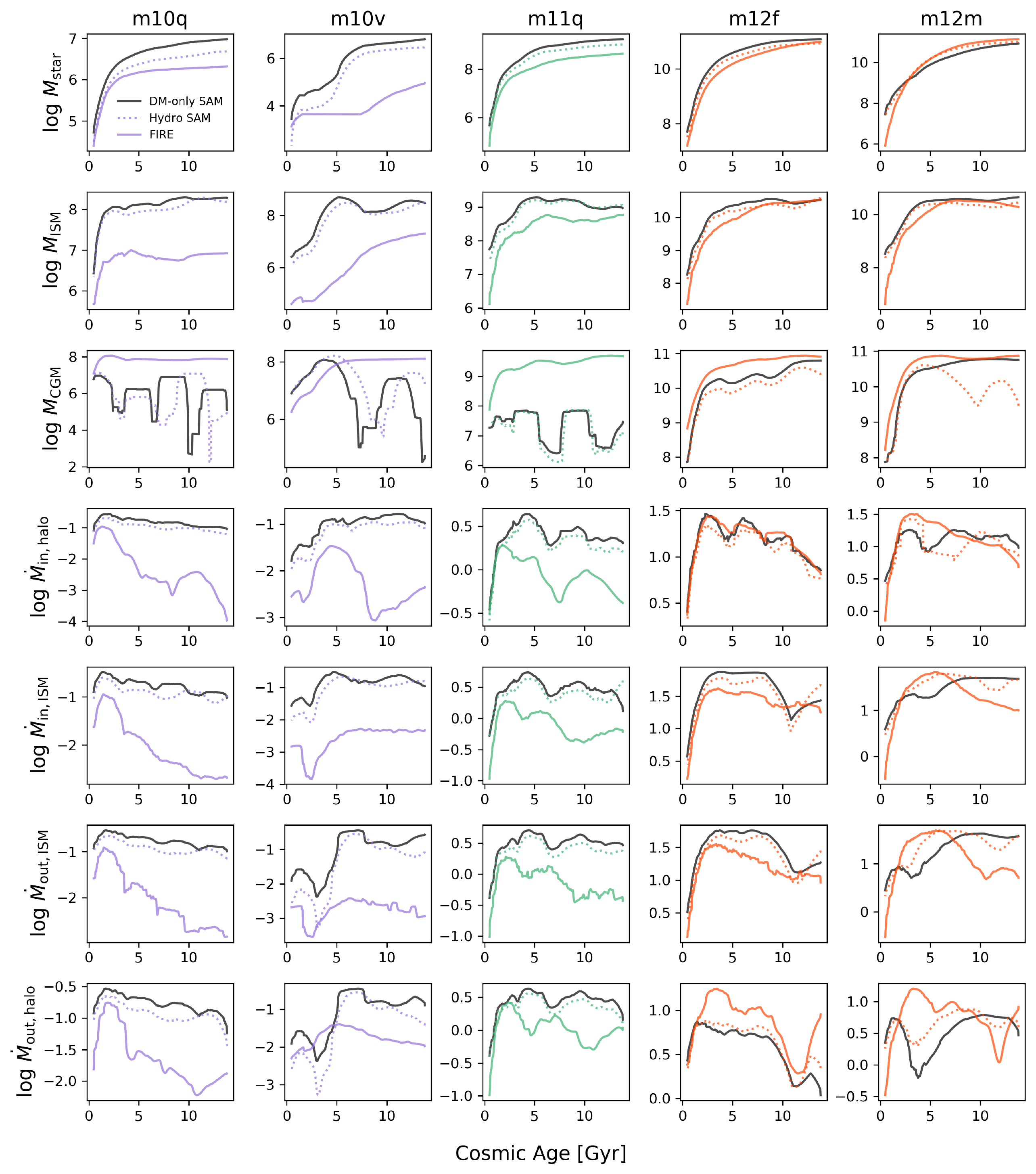}
 \end{center}\caption{Time series of the main properties considered in the paper for the five individual halos with corresponding DM-only simulations. From left to right: m10q, m10v, m11q, m12f and m12m. From top to bottom: stellar mass, ISM mass, CGM mass, halo gas mass inflow rate, ISM mass inflow rate, ISM mass outflow rate and halo mass outflow rate. In every panel, the solid black line shows the prediction of the SAM when run on the DM-only simulation merger trees. The other two curves follow the same convention as the individual halo panels in the figures from the main body of the paper: solid colored curves for the FIRE-2 measurements and dotted colored curves for the SAM predictions using the full hydrodynamical simulation merger trees. The colors show the mass bin that each halo belongs to (purple for m10, green for m11 and red for m12). The main takeaway is that our conclusions do not change if we use the SAM results from the DM-only simulation merger trees: the new SAM predictions agree with our fiducial ones relatively well, and hence the DM-only-based SAM trends relative to FIRE-2 remain qualitatively the same.}
 \label{fig:dmonly_sam}
 \end{figure*}

In \autoref{fig:dmonly_sam} we compare the SAM predictions when run on the FIRE-2 hydrodynamical and DM-only merger trees for the same five halos. It is immediately apparent that the two sets of SAM results agree relatively well with each other. As a consequence, the DM-only-based SAM trends relative to FIRE-2 remain qualitatively the same, and our conclusions would not have changed if we used the DM-only simulation merger trees instead of the fiducial hydrodynamical simulation merger trees. For example, the DM-only SAM still predicts similarly higher halo gas inflow rates for low-mass dwarfs than in FIRE-2 (by $\sim2-3$ orders of magnitude). The main systematic difference between the two sets of SAM predictions is that the DM-only-based SAM predicts somewhat higher stellar masses for dwarfs than the hydrodynamic merger tree-based SAM. This might be due to the higher halo concentrations for dwarfs in the DM-only simulations (no adiabatic expansion due to baryons) leading to smaller predicted disk sizes, which in turn causes gas surface densities and hence higher SFRs. In addition, the $z=0$ CGM masses of the two MW-mass halos also agree better with FIRE-2 using the DM-only-based SAM, although the dwarfs continue to show similarly low CGM masses by orders of magnitude in the SAM compared to FIRE-2. Hence, while there are some relatively minor discrepancies that suggest a deeper look at how the SAM treats baryonic effects on DM halos, in the context of the global baryon cycle that is the main focus of this paper, our conclusions remain the same overall.

\end{document}